\documentclass[twocolumn,superscriptaddress,amssymb,amsmath,nobibnotes,aps,prd,showpacs,nofootinbib]{revtex4-1}
\pdfoutput=1
\usepackage{graphicx}
\usepackage{dcolumn}
\usepackage{bm}
\usepackage{subfigure}
\usepackage{ulem}
\usepackage[table]{xcolor}
\usepackage{amsmath}
\usepackage{fancyvrb}
\usepackage{mathrsfs}
\usepackage{graphicx}  
\graphicspath{{img/}}
\usepackage{dcolumn}   
\usepackage{bm}        
\usepackage{amsmath,amssymb}
\usepackage{hyperref}
\usepackage{color}
\definecolor{purple}{rgb}{0.58,0.0,0.83}
\usepackage{caption}
\usepackage{tabularx}
\usepackage{comment}

\def\beq{\begin{equation}}
\def\eeq{\end{equation}}

\definecolor{mycolor}{rgb}{0.3, 0.0, 0.3}

\definecolor{owngreen}{rgb}{0.0, 0.5, 0.0}

\definecolor{amethyst}{rgb}{0.6, 0.4, 0.8}

\newcommand{\changes}[1]{\textcolor{black}{#1}}
\newcommand{\newchanges}[1]{\textcolor{black}{#1}}
\begin{document}

\title{\changes{Data-driven modeling of} rotation curves with artificial neural networks} 

\author{Gabriela Garcia-Arroyo}
\email{arroyo@icf.unam.mx}
\affiliation{Instituto de Ciencias F\'isicas, Universidad Nacional Aut\'onoma de M\'exico,
62210, Cuernavaca, Morelos, M\'exico.}

\author{Isidro G\'omez-Vargas}
\email{isidro.gomezvargas@unige.ch}
\affiliation{Instituto de Ciencias F\'isicas, Universidad Nacional Aut\'onoma de M\'exico,
62210, Cuernavaca, Morelos, M\'exico.}
\affiliation{Department of Astronomy, University of Geneva, Versoix, 1290, Switzerland.}

\author{J. Alberto V\'azquez}
\email{javazquez@icf.unam.mx}
\affiliation{Instituto de Ciencias F\'isicas, Universidad Nacional Aut\'onoma de M\'exico,
62210, Cuernavaca, Morelos, M\'exico.}

\begin{abstract}
Galactic rotation curves are crucial for understanding the distribution of mass in galaxies. Despite advances in precision observations, there are discrepancies between the inferred mass from luminosity and the observed rotational velocities, often attributed to dark matter. While traditional parametric models provide valuable insights, they struggle with complex galactic features like prominent bulges and non-circular motions. In this study, we apply artificial neural networks to \newchanges{generate robust, data-driven models, tailored to each galaxy,} for the rotation curves of spiral galaxies using high-quality observational data. Our approach demonstrates that neural networks can effectively capture the intricate structure of rotation curves without relying on predefined astrophysical assumptions. By comparing the \changes{data-based models} with the Navarro-Frenk-White model under two different assumptions for the stellar component, we \changes{classify} galaxies based on the model that best fits their rotation curves, offering insights into the limitations and strengths of both \changes{theoretical and data-based} methods.  This work highlights the potential of machine learning techniques \changes{ in identifying galaxies whose dynamics are not well captured by standard theoretical models, pointing to the need for more refined physical descriptions.}

\end{abstract}
\maketitle

\section{Introduction}
\label{sec:intro}

Since the pioneering measurements of galactic rotation curves (RCs) \citep{1914LowOB...2...66S}, to the first precision observations \citep{Rubin:1970zza}, they have been an indispensable tool for determining the distribution of mass in galaxies. Although rotation curves of different types of galaxies exhibit their own distinct features, they share several common characteristics \citep{Karukes:2017kne}. Notably, in most RCs the matter distribution inferred from luminosity shows a discrepancy with that derived from the observed rotation velocities. An explanation for this inconsistency is that galaxies are embedded in extended Dark Matter (DM) halos, whose gravitational influence is appreciated in the rotation velocities, not in the luminosity. 

Galaxies can be classified into Low-Surface-Brightness (LSB) or High-Surface-Brightness (HSB) types based on their surface brightness, irrespective of morphology. This classification has implications for their associated RCs.
LSB galaxies contain more gas than HSB galaxies. However, they exhibit lower surface brightness due to their very low star formation rates, and they are DM-dominated at all radii, including their centers \citep{deBlok:1997zlw, Pahwa_2018}.
On the other hand, in addition to the gas component, the stellar component can significantly contribute to the dynamics of high surface brightness galaxies, particularly in the inner regions. Consequently, the DM halo parameters become sensitive to the assumed mass-to-light ratio of the stellar disk, $\Upsilon_{\star}$, which is used to convert the luminous mass into gravitational mass \citep{Sellwood:1998kj}. \changes{ RCs in HSB galaxies may show a faster increase in rotational velocities towards the center, indicative of higher mass concentration.}

Measurements of RCs in LSB and HSB galaxies present distinct challenges and considerations. For example, LSB galaxies require specialized techniques and longer exposure times for detection, particularly in their faint outer regions. Higher spatial resolution is crucial for studying the fine details of LSB galaxies \citep{deBlok:2002vgq}. In contrast, HSB galaxies are generally easier to detect, and lower spatial resolution suffices for central studies. 

The inherent complexity of galactic dynamics, whether LSB or HSB, coupled with the limitations of observational methods, present challenges in obtaining precise and detailed RCs data. Even with perfect observations, obtaining sufficient information to construct a comprehensive theoretical model that fully captures the mass distribution within these objects remains a significant challenge. \changes{Therefore, the simplifying assumptions about the mass distribution in current theoretical models limit their ability to reproduce the full dynamical complexity observed in galaxies.} In response to these challenges, the use of various statistical and machine learning methods is increasing, offering valuable new tools for astrophysical data analysis \citep{fernandez2019galaxy, yi2022automatic, thuruthipilly2024shedding, tanoglidis2021deepshadows, cortes2024galaxies}. In particular, machine learning methods have already been used in RCs data for curve fitting \citep{arguelles2023galaxy}, 
and to model the contribution and density profile of DM \citep{Pato:2015tja}.

In \citet{fernandez2019galaxy}, the authors present an innovative approach to modeling rotation curves using a non-parametric reconstruction based on Locally Weighted Scatterplot Smoothing (LOESS) and Simulation and Extrapolation (SIMEX) regression. \changes{This methodology extracts information directly from observational data with minimal prior assumptions about the underlying astrophysical model. \changes{Similar approaches have been developed for model-independent data analysis of astrophysical and cosmological observations}, utilizing methods such as Principal Component Analysis \citep{sharma2020reconstruction, Escamilla:2021uoj}, Gaussian processes \citep{Velazquez:2024aya, Keeley:2020aym, l2020defying, escamilla2023model}, and, more recently, artificial neural networks \citep{wang2020reconstructing, gomez2023neuralrecs}. These approaches share a common goal: to infer empirical relationships directly from data while minimizing dependence on explicit physical parameterizations.}

\changes{In this paper, we apply a data-driven artificial neural network (ANN) framework that integrates Monte Carlo dropout for regularization and uncertainty quantification, together with genetic algorithms for optimizing the network hyperparameters. \newchanges{These techniques are applied independently to each galaxy, yielding tailored data-driven models that capture the specific characteristics of each rotation curve.} This pipeline, previously developed and validated in other cosmological contexts \cite{gomez2023neuralrecs, gomez2023neuralgenetic, mitragomez2024dark}, is here employed to model the rotation curves of spiral galaxies. Our approach is conceptually similar to that of \citet{fernandez2019galaxy}, who used the LOESS+SIMEX interpolation method for dark-matter-dominated galaxies. However, unlike LOESS, which assumes a local polynomial function while SIMEX applies a correction for bias, introduced by measurement errors in the data, our neural network framework does not rely on any predefined functional form. While LOESS+SIMEX performs well for smooth data, it remains limited by the choice of polynomial degree, smoothing parameters, and its inherently local nature, fitting the data piecewise rather than modeling the whole set of data points simultaneously. In contrast, ANNs learn a global, data-driven mapping that represents the full rotation-curve behavior of each galaxy by simultaneously modeling the three quantities present in the observational RCs datasets: radius, rotational velocity, and statistical uncertainty, treating the radius as the independent variable and the other two as coupled dependent outputs. This distinction clarifies that ANNs do not merely interpolate the data but instead learn the underlying functional relationships, allowing them to model complex, nonlinear, and irregular velocity structures without physical prior assumptions.}

\changes{In this work, } we consider 17 observed rotation curves of spiral galaxies from the high-quality data of The HI Nearby Galaxy Survey (THINGS) \citep{Walter:1985psa,deBlok:2008wp}. These galaxies are mostly spiral HSB galaxies whose mass profile can be described by an empirical DM density profile \citep{Einasto, Navarro:1995iw, Burkert:1995yz, Begeman:1991iy, Salucci:2007tm, Mastache:2012ep, Navarro-Boullosa:2023bya}, plus a stellar distribution along with a gas component. We recommend \cite{Sofue:2000jx} for a historical context of this survey. 

The mass models for \changes{the THINGS galaxies} have been constructed by combining their HI RCs \citep{deBlok:2008wp} with information about the gas and stellar distribution within them, the latter obtained from $3.6 \mu $m data from the Spitzer Infrared Nearby Galaxies Survey (SINGS) \citep{Kennicutt:2003dc}. Through this combination, the value of $\Upsilon_{\star}$ is derived. Then, the RC associated with the DM halo is obtained and fitted using a parametric density model, \changes{such that, in practice, with this procedure the total parametric RC follows directly from the observed baryons and the total measured velocity field, whereas the ANN model does not rely on any assumed $\Upsilon_{\star}$ or density halo profile. }

\changes{For comparison purposes, we adopt a Navarro-Frenk-White (NFW)} dark matter halo as a theoretical model, along with two different assumptions for $\Upsilon_{\star}$ (fixed and free), we carry out
a detailed comparison of the mean squared error (MSE) between the theoretical models and the ANN reconstructions, \changes{and group the} galaxies into three categories: those where the ANN performs better, those where it outperforms the fixed $\Upsilon_{\star}$ model but not the free $\Upsilon_{\star}$ model, and those where both parametric models outperform the ANN. This classification reveals important physical insights into the dynamics of galaxies and demonstrates the potential of ANN-based methods. \changes{In this sense, our results highlight how non-parametric methods can probe the interplay between baryonic and dark matter distributions without relying on predefined astrophysical assumptions.}
\\

This paper is organized as follows. In Sec.~\ref{sec:background}, we present the statistical basics to understand the method we apply in this manuscript. 
Sec.~\ref{sec:galaxy} describes the sample of galaxies, providing specific details about their distribution profiles.
In Sec.~\ref{sec:methodology}, we describe the methodology followed in this work to accurately reconstruct the RCs for the data sets employed, while
in Sec.~\ref{sec:results1}, we show the main results of our reconstructions. 
In addition, we compare the performance of our approach with that of traditional parametric models.
In Sec.~\ref{sec:conclusions} we give some final comments and conclusions.

\section{Deep Learning Framework}
\label{sec:background}



\changes{Deep learning provides a powerful framework for generating data-driven models and capturing complex, non-linear relationships directly from observations. It is a field of machine learning that centers on the design and training of artificial neural networks to learn patterns and representations from data. Rather than relying on explicit analytical forms or astrophysical prescriptions, deep learning methods extract the underlying structure of the data itself, offering a flexible approach for modeling galaxy rotation curves.}

Artificial neural networks have the theoretical ability to approximate any continuous non-linear function \citep{hornik1990universal} and have been successfully applied to a wide variety of scientific problems, including regression, classification, and generative modeling. Their flexibility and capacity to represent complex data distributions make them particularly well-suited to astrophysical analyzes where the physical relationships are not fully constrained. For general overviews of neural network theory and applications, we refer the reader to \citet{nielsen2015neural, goodfellow2016deep}.
\\

In this work, we employ feed-forward neural networks to model the rotation curves of spiral galaxies directly from observational data. The networks take the radial distance as input and predict the corresponding rotational velocity and its associated uncertainty, enabling a fully data-driven reconstruction of the empirical relationship 
$V(r)$.

\changes{To enhance the reliability of our results, we adopt the framework proposed by \citet{gomez2023neuralrecs, gomez2023neuralgenetic, mitragomez2024dark}, which incorporates two complementary deep learning components:}

\begin{itemize}
    \item \textbf{Monte Carlo Dropout (MC-DO)}. A probabilistic regularization method that estimates predictive uncertainties by performing multiple stochastic forward passes through the network \citep{gal2015dropout, benatan2023enhancing}. This approach provides an efficient means of quantifying the uncertainty of the model while mitigating overfitting.

    \item \textbf{Genetic Algorithms (GAs).} A global optimization strategy used to automatically determine the hyperparameters of the neural network \citep{gomez2023neuralgenetic}. By exploring a broad configuration space of possible architectures, GAs improve convergence and ensure robust model performance.

\end{itemize}

\newchanges{In our method, MC-DO and GAs play complementary roles. MC-DO acts as a regularizer during training; keeping dropout active during inference and performing repeated stochastic forward passes yields a predictive distribution whose variance reflects epistemic uncertainty, while the mean provides a robust prediction. The GA performs a global search over network architectures and training hyperparameters that control model capacity, such as the number of hidden layers, the number of neurons per layer, the batch size, and the learning rate. Because each candidate configuration is evaluated based on its performance on validation data, the selected model is the one that best generalizes instead of the one that best fits the training set. Together, these components facilitate a proper balance between bias and variance: MC-DO avoids overfitting within a given architecture, while GA selects a level of complexity justified by the validation performance. In this way, our deep learning framework can balance flexibility, interpretability, and statistical rigor, enabling us to model galactic rotation curves directly from the data while providing meaningful uncertainty estimates. }

\section{Galaxy sample for a parametric description}
\label{sec:galaxy}


The observed rotation curves in the THINGS sample correspond to circular orbits in the equatorial plane and, in general, have contributions from three different mass distributions: gas $(V_{\rm gas})$, stars $(V_{\star})$, and a DM halo $(V_{\rm halo})$.  
The individual velocities of the stars and gas, calculated from the observed baryonic mass within the galaxy, represent the speeds that each of these components would induce on a test particle if they were isolated and unaffected by external influences. On the other hand, the halo contribution is introduced to fit the total observed velocity.

These individual contributions are taken into account in the calculation of the total parametric RC as follows: 
\begin{equation}\label{eq:total_velocity}
    V^2 = V^{2}_{\rm gas}+ \Upsilon_{\star}V^{2}_{\star}+V^{2}_{\rm halo}\, ,
\end{equation}
\changes{where $\Upsilon_{\star}$ modulates the stellar contribution to the total velocity}; $V_{\star}$ has units km/s$\left(L_{\odot}/M_{\odot}\right)^{1/2}$. Theoretically, this is a parametric equation with parameters associated with specific models of the stellar disk and the DM model. 
Considering that the main goal of this paper is to compare the fit provided by theoretical models with non-parametric reconstructions, we use the theoretical parametric results from Tables 3 and 6 of \citet{deBlok:2008wp}. Both models assume an NFW distribution for the dark matter halo but with different assumptions regarding $\Upsilon_{\star}$. In the following subsections, we outline these assumptions related to each mass component on which the utilized data is based. More specific details on galaxies can be found in \cite{deBlok:2008wp, Walter:1985psa}.

\subsection{Gas distribution}

The neutral gas is one of the baryonic components that contribute to the total mass of a galaxy, and the circular velocity that it induces on a test particle is required as an input to Eq. \ref{eq:total_velocity}. For this sample of galaxies, the rotation curve was derived in 
\cite{Walter:1985psa} using integrated hydrogen (HI) emission maps to initially obtain the surface density profiles, which were then corrected by a factor of $1.4$ to take into account the contribution of helium and metals. Finally, the corresponding rotation curve, $V_{\rm gas}$, was calculated assuming an isolated thin disk distribution. 

\subsection{Stellar distribution}
The circular velocity derived from this component, $\Upsilon_{\star}V^{2}_{\star}$, which will be an input of Eq. \ref{eq:total_velocity}, requires that one previously know its mass distribution. However, the stellar density profile is not directly measured; measurements are based on the observed surface brightness profile, $I(r)$.  
In some galaxies within the sample, this component can be further decomposed into central (bulge, inner) and  outer brightness profiles, each one considered to reside in an isolated thick disk, whose radial distribution is characterized by an exponential profile \citep{Freeman:1970mx}:
\begin{equation}\label{eq:exponential_profile}
   \Sigma(r)=\Upsilon_{\star}^{3.6}I_{0}^{3.6}\exp^{-r/R_{\rm d}}\, ,
\end{equation}
where $I_0$ is the surface brightness at $r=0$ and $R_{\rm d}$ is the characteristic scale of the bulge/outer component\footnote{If both components are present, the associated circular velocity should be: $\Upsilon_{\star}V_{\star}^2=\Upsilon_{\star}^{\text{bulge}}V_{\star, \text{bulge}}^2+\Upsilon_{\star}^{\text{outer}}V_{\star, \text{outer}}^2$, reflecting two luminosity profiles, each one with its parameters.
}. Observationally, it has been determined that the distribution along the $z$ axis follows a $\text{sech}^{2}(z/2z_d)$, with $z_d=R_d/5$, such that the volumetric density is the separable product of $\Sigma(r)$ and $\text{sech}^{2}(z/2z_d)$.
This is the assumption taken for the THINGS galaxies, whose parameter values can be found in \cite{Oh:2008ww} and were determined by observations in the $3.6\mu \rm{m}$ band,
assuming an initial mass function (IMF) of scaled-down Salpeter, also known as `diet Salpeter' \citep{Bell:2000jt}. This model assumption is referred to as $\Upsilon_{\star}^{\rm fix}$ in this paper.

 Although $\Upsilon_{\star}^{\rm fix}$ is determined by the observed brightness profile, it carries the biggest uncertainties in the modeling of baryonic components. Therefore, in \cite{deBlok:2008wp}, the authors consider an alternative model in which the mass-to-light ratio is left as a free parameter through the relation $\Upsilon_{\star}^{\rm free}=f^{D}\Upsilon_{\star}^{\rm fix}$, where $f^D$ has to be obtained assuming a particular DM distribution to fit the total RC, as implied by Eq~\ref{eq:total_velocity}. In this paper, we refer to this assumption as $\Upsilon_{\star}^{\rm free}$.

\subsection{Dark matter distribution}

According to the standard cosmological model, we assume the dark matter halo distribution with the NFW profile ~\citep{Navarro:1995iw}, 
featuring  the characteristic density $\rho_i$ (which is related to the density at the time of the halo collapse), and a characteristic scale, $R_s$:
\begin{equation}
    \rho_{\rm NFW}(r)=\frac{\rho_i}{\left(r/R_{s}\right)\left(1+r/R_{s}\right)^2}\ .
    \end{equation}
Generally, the two free parameters $\rho_i$ and $R_s$ are used to define new parameters, for example, the radius $r_{200}$, at which the enclosed density is 200 times the critical density. In addition, a concentration parameter $c=r_{200}/R_{s}$ is introduced, and the reference velocity $V_{200}$ is measured at $r_{200}$. 
Consequently, the new set of free parameters for this parametric model becomes $c$ and $V_{200}$. For the galaxies analyzed in this work, these values have been extracted from Tables 3 and 6 of \citet{deBlok:2008wp}. 


   

\begin{figure}[t!]
 \captionsetup{justification=raggedright,singlelinecheck=false,font=footnotesize}
 \centering
    \includegraphics[width=0.5\textwidth]{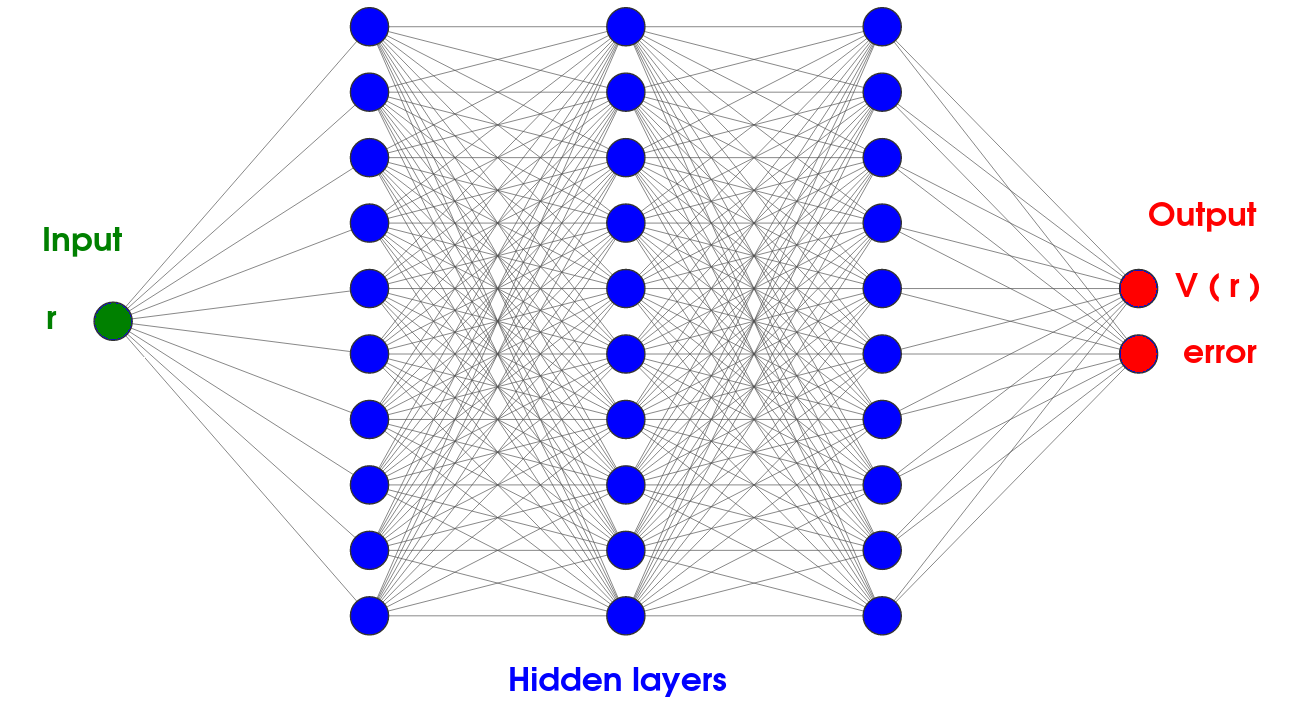}
    \caption{ANNs general architecture for the rotation curves. The number of hidden layers and their number of neurons depend on each galaxy because for each case a different neural network model is generated. In all the cases, the input is the values of the radius, and the output is the velocity and the statistical error.}
    \label{fig:archann}
\end{figure}

\begin{table}
 \captionsetup{justification=raggedright,singlelinecheck=false,font=footnotesize}
    \centering
    \scriptsize
    \begin{tabular}{|c c c c|}
    \hline
    Galaxy   & \# hidden layers & \# neurons & batch size  \\
    \hline
    NGC4736 & 4 & 200 & 8 \\
    \hline
    NGC2841 & 3 & 150 & 8 \\
    \hline
    NGC925. & 4 & 200 & 4 \\
    \hline
    NGC2903 & 4 & 150 & 8 \\
    \hline
    DDO154. & 4 & 200 & 4 \\
    \hline
    NGC6946 & 4 & 200 & 8 \\
    \hline
    NGC7793 & 4 & 150 & 4 \\
    \hline
    NGC2976 & 4 & 200 & 4 \\
    \hline
    NGC7331 & 3 & 200 & 4 \\
    \hline
    NGC3198 & 3 & 150 & 4 \\
    \hline
    NGC3621 & 4 & 200 & 4 \\
    \hline
    NGC3521 & 3 & 200 & 4 \\
    \hline
    NGC5055 & 4 & 100 & 4 \\
    \hline
    NGC2366 & 4 & 200 & 8 \\
    \hline
    NGC2403 & 3 & 150 & 8 \\
    \hline
    NGC3031 & 3 & 200 & 4 \\
    \hline
    IC2574. & 3 & 200 & 4 \\
    \hline
    \end{tabular}
    \caption{Best hyperparameter combinations for the 17 ANN models founded by genetic algorithms, each one corresponding to a different galaxy.}
    \label{tab:hyperparameters_reordered}
\end{table}

\section{Methodology}\label{sec:methodology}

Throughout this work, we employ data-driven modeling with artificial neural networks to reconstruct rotation curves. As discussed in the Introduction and Section \ref{sec:background}, this approach enables the generation of computational models directly from the data, without relying on specific astrophysical or statistical assumptions. However, the accuracy and reliability of such models are inherently constrained by the quality and range of the training data, in our case, the observed radii and velocities. Therefore, careful training and regularization of the networks are essential to ensure robust and physically meaningful reconstructions.

We follow the methodology proposed in \citet{gomez2023neuralrecs}, combined with the hyperparameter optimization scheme based on genetic algorithms developed in \citet{gomez2023neuralgenetic}. In this framework, the ANN receives the radius as input and returns two outputs: the rotational velocity $V(r)$ and its associated statistical error (Figure~\ref{fig:archann}). The use of Monte Carlo dropout (MCDO) introduces a Bayesian behavior in the predictions, providing uncertainty estimates for each prediction \citep{gal2015dropout, benatan2023enhancing}. \changes{While conventional Machine Learning models do not provide intrinsic uncertainty estimates—the loss function only quantifies the discrepancy between predictions and observations and does not capture epistemic uncertainty, the MCDO technique enables an estimation of predictive variance. By keeping dropout active during inference and performing multiple stochastic forward passes, we obtain both a mean prediction (the expected output) and a variance that reflects the confidence of the model. This procedure yields more reliable predictions, particularly in regions where the observational data are sparse or noisy.}

\changes{Given the relatively small number of observational points available for each galaxy’s rotation curve (typically on the order of hundreds), our primary objective is to balance model accuracy with generalization capability while avoiding overfitting. Traditional interpolation methods might fit the data more tightly but lack the ability to capture underlying physical trends. By contrast, our framework employs both MCDO and GAs to ensure that the networks are flexible yet well-trained, with a balance between bias and variance, and without underfitting or overfitting.}

After initial exploration of the data and preliminary training, we define the hyperparameter space with the number of hidden layers $\in [3, 4]$, neurons per layer $\in [50, 100, 150, 200]$, and batch size $\in [4, 8, 16, 32]$. We then use the \texttt{nnogada} code \citep{gomez2023neuralgenetic}, based on a genetic algorithm from the DEAP library \citep{DEAP_JMLR2012}, to optimize the neural network architecture for each galaxy. \newchanges{Candidate configurations are evaluated using the mean squared error in the validation set, such that the selected architecture balances model complexity and generalization, mitigating both underfitting and overfitting.} The resulting neural network architectures are listed in Table~\ref{tab:hyperparameters_reordered}.

For training, we used Adam optimizer \citep{kingma2014adam}with a fixed learning rate of $5\times10^{-4}$, ReLU activation functions in hidden layers, and linear activation in the output layer. All models were trained for 1000 epochs. As in \citet{gomez2023neuralrecs}, we used the Monte Carlo dropout method \citep{gal2015dropout} with a fixed dropout rate of 0.3 to ensure robust models. \changes{The learning rate and number of epochs were fixed rather than included in the GA search to ensure comparability and stability across different galaxies. Including them in the optimization would increase the search space. Moreover, extensive testing showed that a learning rate of $5\times10^{-4}$ provided a good balance between convergence speed and stability, while 1000 epochs were sufficient for all models to reach a plateau in validation loss without signs of overfitting. Thus, fixing these values ensures consistent training behavior and allows the GA to focus on structural hyperparameters that most strongly affect model complexity and generalization.}

Once the neural networks were trained, we performed predictions across different radii within the range of the original rotation curve data, yielding a \changes{data-based} model of $V(r)$. These reconstructions incorporated both the statistical errors of the original data and the uncertainty in the predictions of the neural network. The results of these reconstructions are discussed in the next section.

To compare neural network reconstructions with theoretical parametric models (NFW fixed and free), we used the mean square error (MSE) as the evaluation metric:
\begin{equation} \text{MSE} = \frac{1}{n} \sum_{i=1}^{n} (y_i - \hat{y}_i)^2, \end{equation}
where $n$ is the number of samples, $y_i$ represents the observed value, and $\hat{y}_i$ is the predicted value. This comparison allowed to classify galaxies based on which model (ANN, NFW fixed, or NFW free) provided the best fit to the observed data. The MSE-based classification helps identify trends in the data and analyze the strengths and limitations of both parametric and \changes{neural network} models, offering insights into the complex mass distribution and galaxy dynamics. 

For complete reproducibility, all training notebooks, reconstruction plots, and hyperparameter configurations are publicly available on GitHub\footnote{\url{https://github.com/igomezv/Reconstructing-RC-with-ANN}}.

\begin{table*}
    \captionsetup{justification=raggedright,singlelinecheck=false,font=footnotesize}
    \centering
    \scriptsize
    \begin{tabular}{|c|c|c|c|}
    \hline
    Galaxy & MSE(fix, data) & MSE(free, data) & MSE(neural, data)\\
    \hline
    \hline
    DDO154 & 2.83 & 2.91 & 3.50 \\ 
    \hline
    NGC2903 & 10.04 & 6.85 & 17.38  \\ 
    \hline
    NGC3621 & 12.24 & 10.17 & 22.10  \\ 
    \hline
    NGC2403 & 11.56 & 11.35 & 12.34  \\ 
    \hline
    NGC3031 & 144.49 & 103.98 & 105.42 \\ 
    \hline
    NGC2841 & 32.43 & 21.50 & 27.68  \\ 
    \hline
    NGC3198 & 100.39 & 7.98 & 26.03  \\ 
    \hline
    NGC5055 & 239.95 & 42.84 & 124.15  \\ 
    \hline
    NGC4736 & 46.66 & 43.68 & 42.66  \\ 
    \hline
    NGC925 & 66.52 & 34.65 & 25.11  \\ 
    \hline
    NGC6946 & 119.78 & 51.66 & 26.71  \\ 
    \hline
    NGC7793 & 59.74 & 53.47 & 21.39 \\ 
    \hline
    NGC2976 & 25.88 & 11.15 & 3.10  \\ 
    \hline
    NGC7331 & 309.23 & 19.78 & 12.64  \\ 
    \hline
    NGC3521 & 335.01 & 225.37 & 62.68 \\ 
    \hline
    NGC2366 & 15.75 & 7.60 & 2.97  \\ 
    \hline
    IC2574 & 14.59 & 8.55 & 2.47 \\ 
    \hline
     Means & 91.01 & 39.03 & 31.67  \\
     \hline
    \end{tabular}
    \caption{\changes{Mean squared errors.} \textit{First column:} Between the NFW with $\Upsilon_{\star}^{\rm fix}$ versus the observational data. \textit{Second column:} NFW with $\Upsilon_{\star}^{\rm free}$  versus the observational data. \textit{Third column:} Neural network reconstruction versus the observational data.}
    \label{tab:results}
\end{table*}

\section{Results}
\label{sec:results1}

  \begin{figure*}
    \centering    \captionsetup{justification=raggedright,singlelinecheck=false,font=footnotesize}
    \makebox[10.cm][c]{
            \includegraphics[trim=3mm 0mm 0.0mm 0mm, clip, width=6.0cm, height=4.1cm]{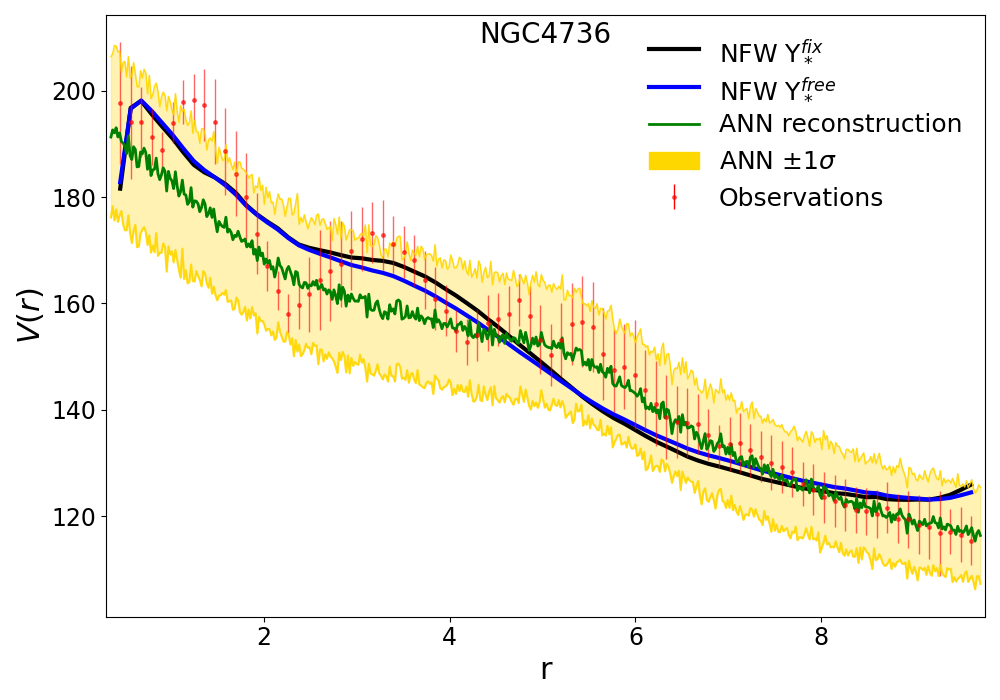}
            \includegraphics[trim=3mm 0mm 0.0mm 0mm, clip, width=6.0cm, height=4.1cm]{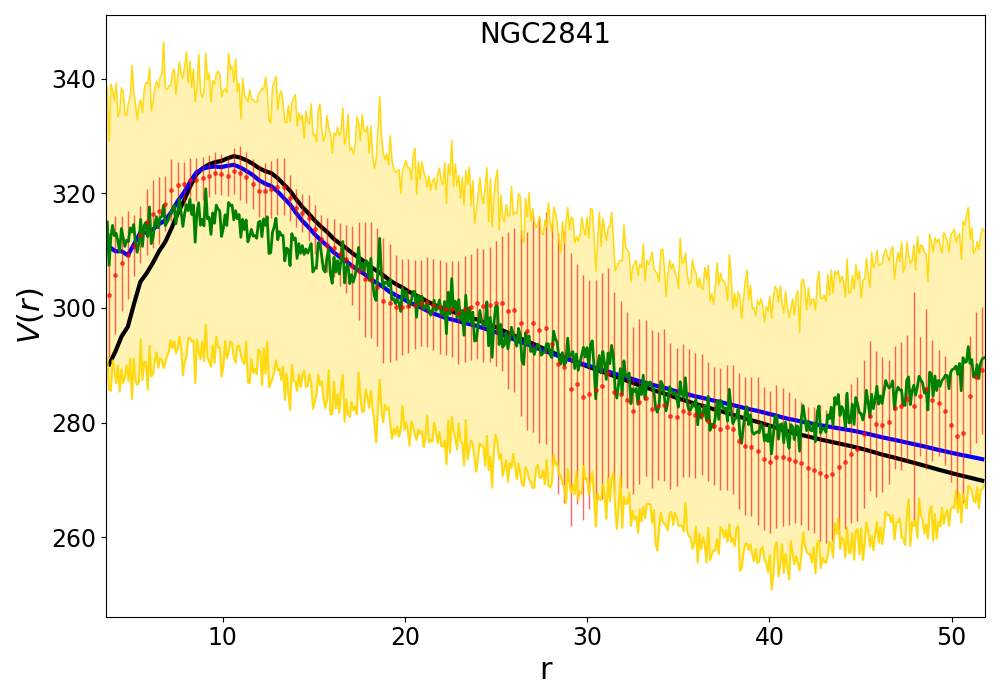}
            \includegraphics[trim=3mm 0mm 0.0mm 0mm, clip, width=6.0cm, height=4.1cm]{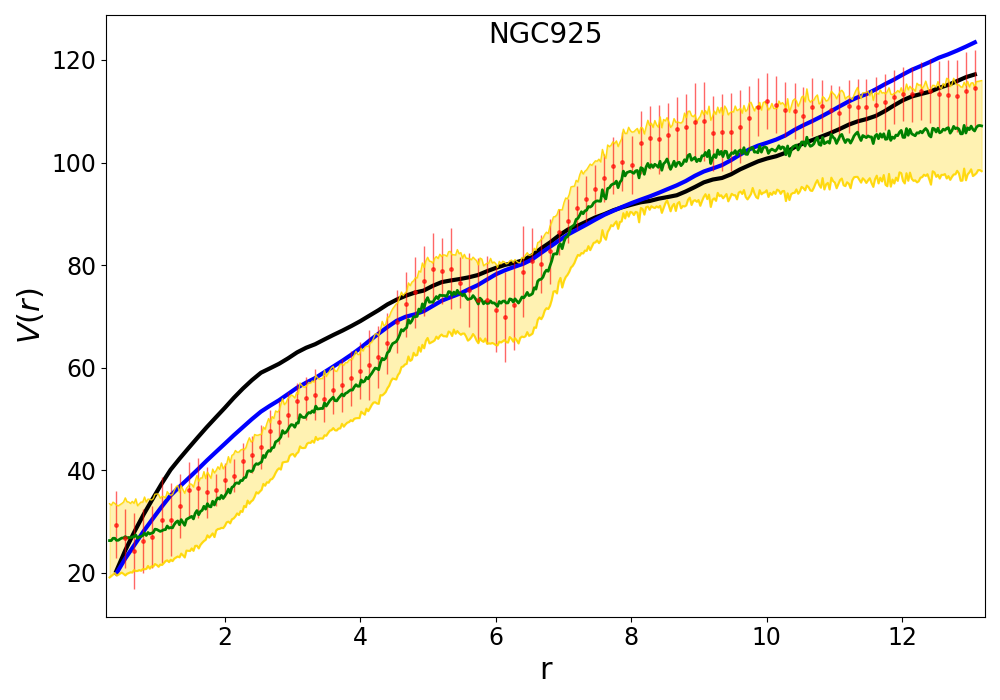}
        }
        \makebox[10.cm][c]{
            \includegraphics[trim=3mm 0mm 0.0mm 0mm, clip, width=6.0cm, height=4.1cm]{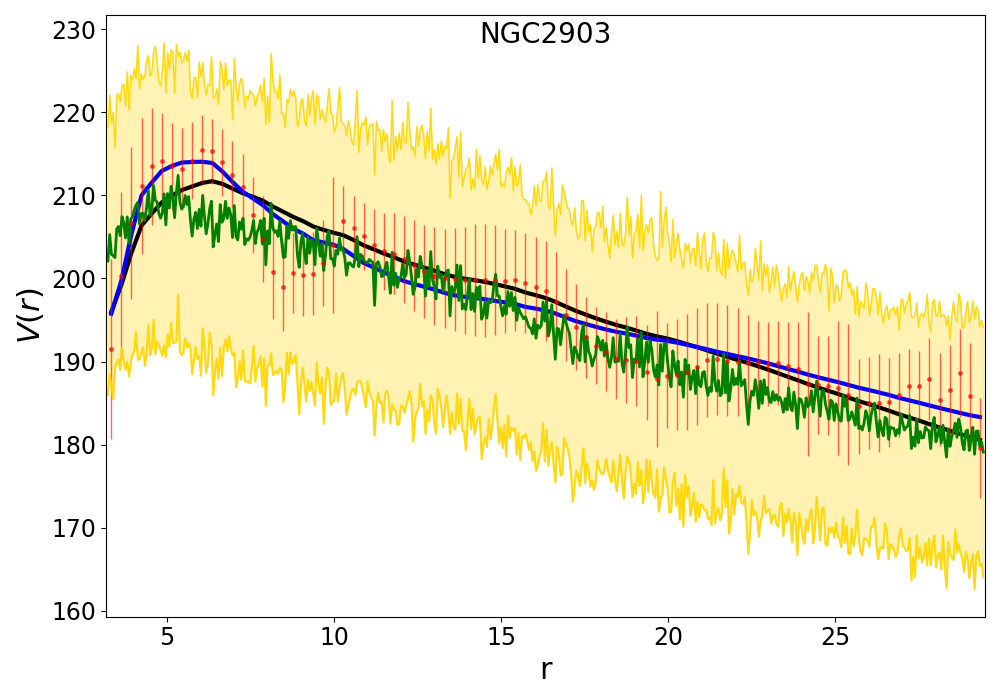}
            \includegraphics[trim=3mm 0mm 0.0mm 0mm, clip, width=6.0cm, height=4.1cm]{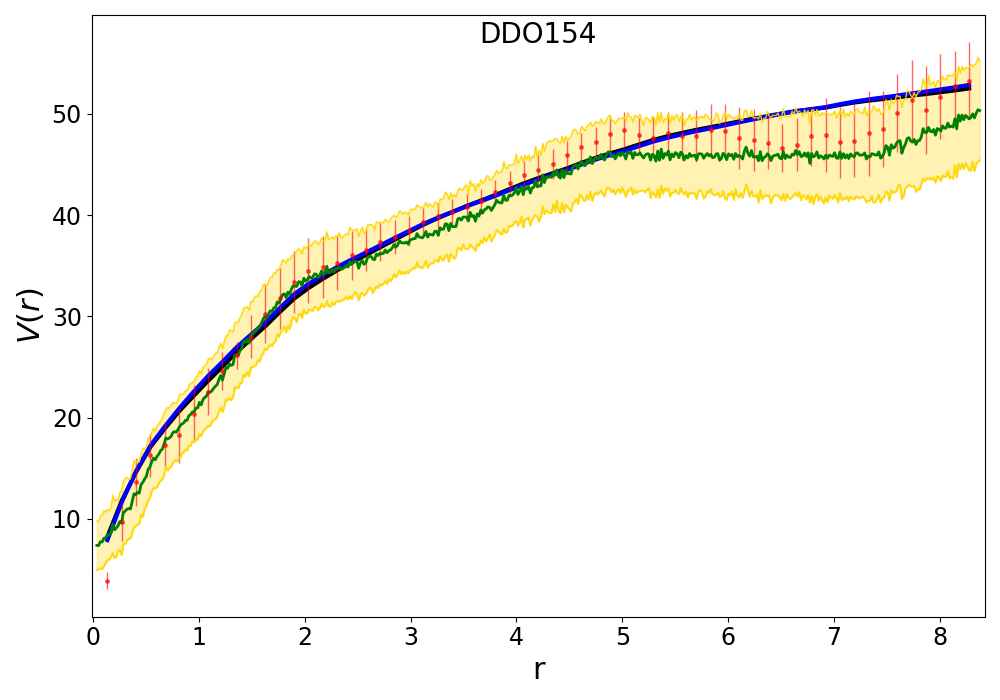} 
            \includegraphics[trim=3mm 0mm 0.0mm 0mm, clip, width=6.0cm, height=4.1cm]{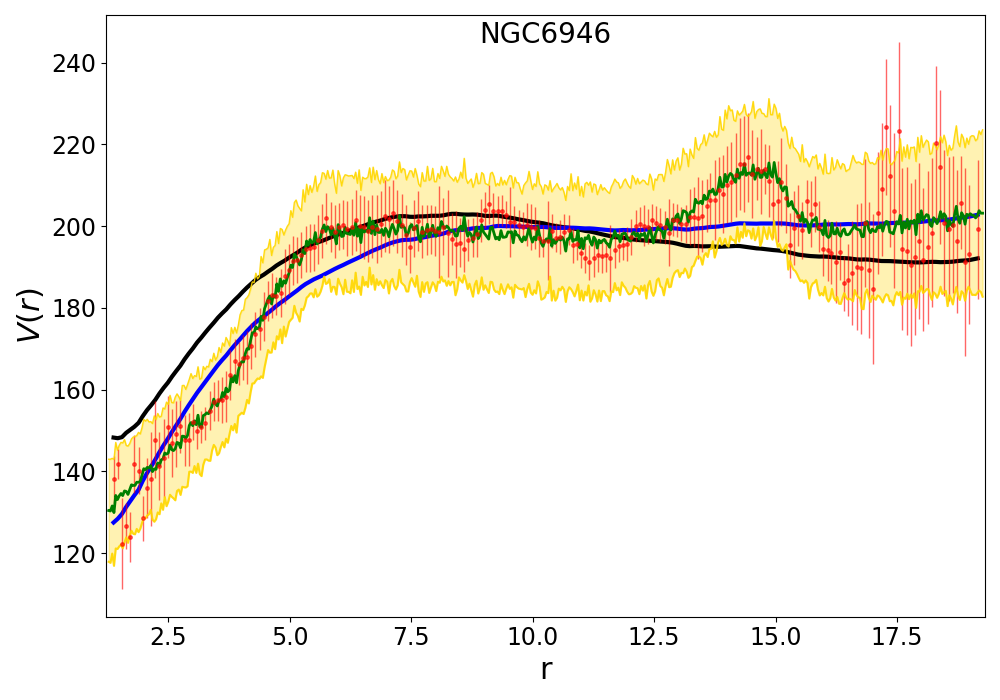} 
        }
        \makebox[10.cm][c]{
            \includegraphics[trim=3mm 0mm 0.0mm 0mm, clip, width=6.0cm, height=4.1cm]{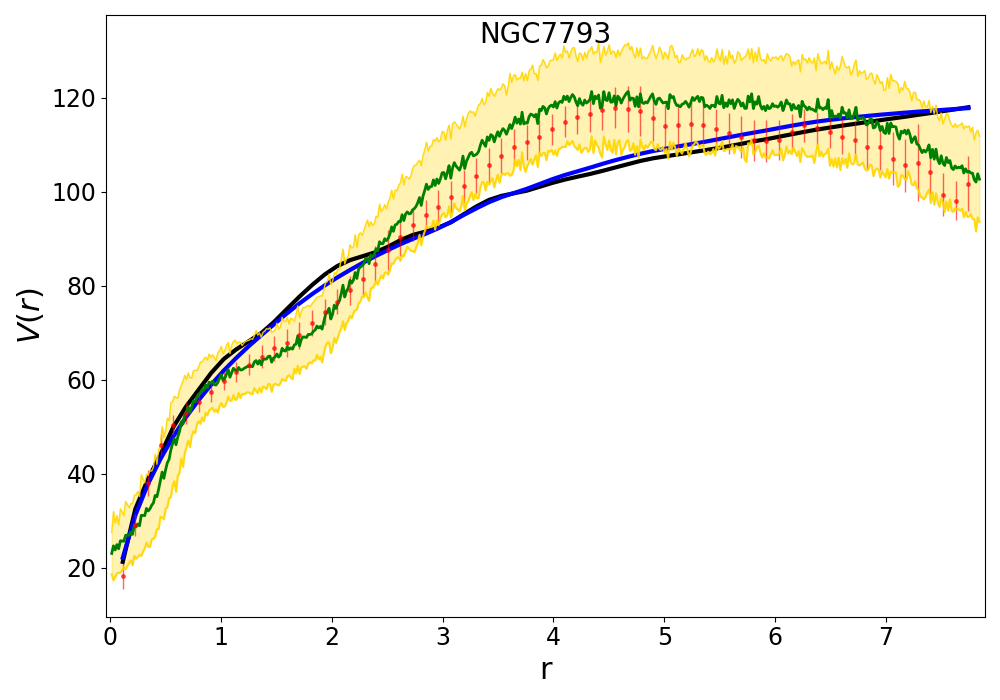}
            \includegraphics[trim=3mm 0mm 0.0mm 0mm, clip, width=6.0cm, height=4.1cm]{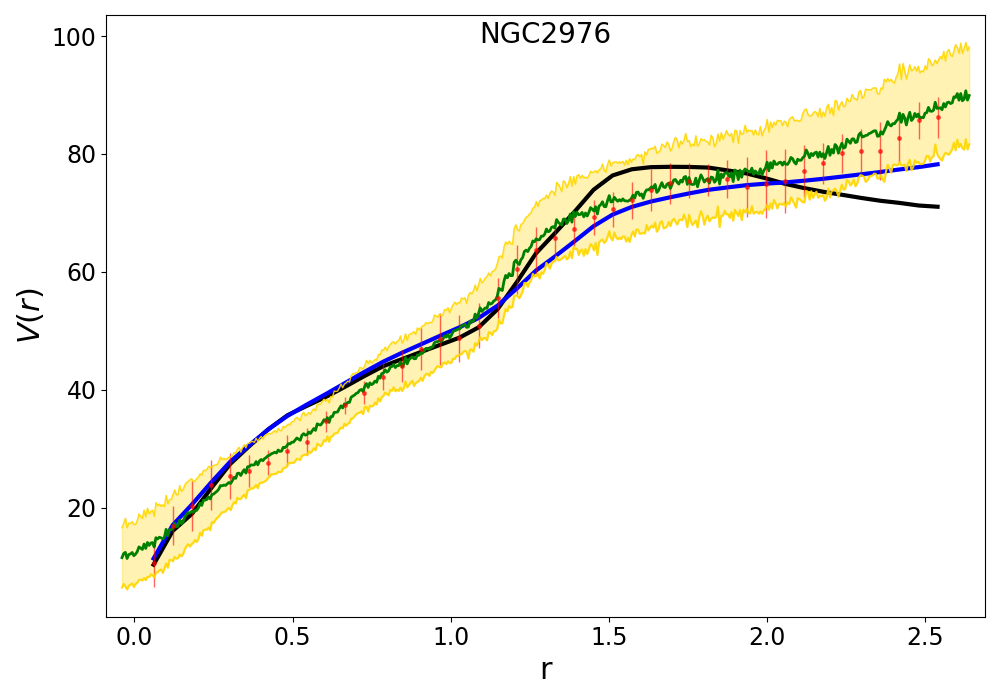}         
            \includegraphics[trim=3mm 0mm 0.0mm 0mm, clip, width=6.0cm, height=4.1cm]{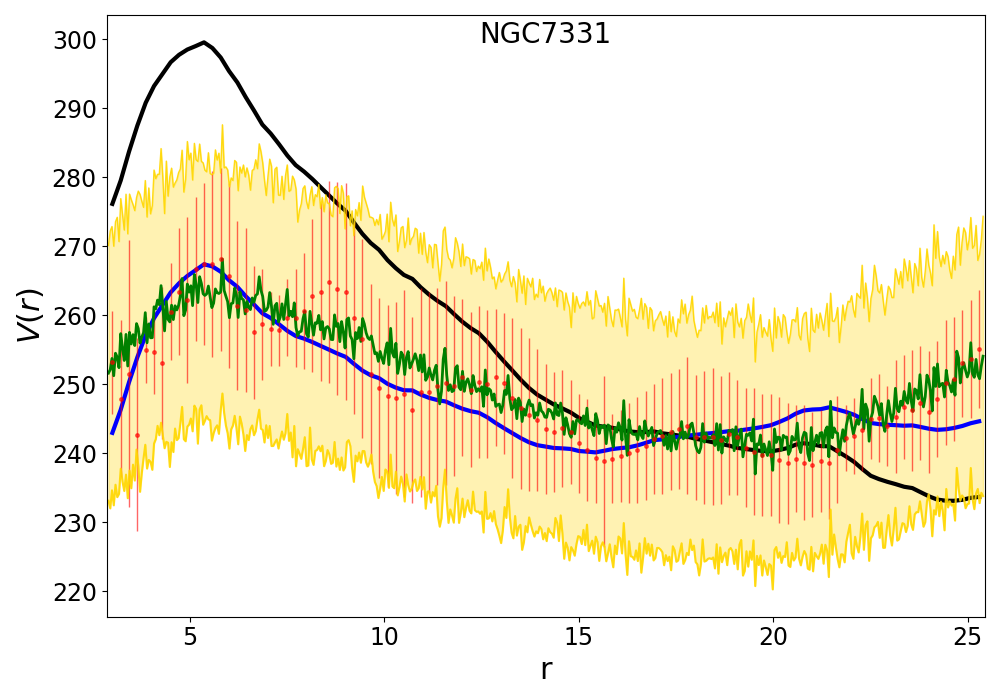} 
        }
        \makebox[10.cm][c]{
            \includegraphics[trim=3mm 0mm 0.0mm 0mm, clip, width=6.0cm, height=4.1cm]{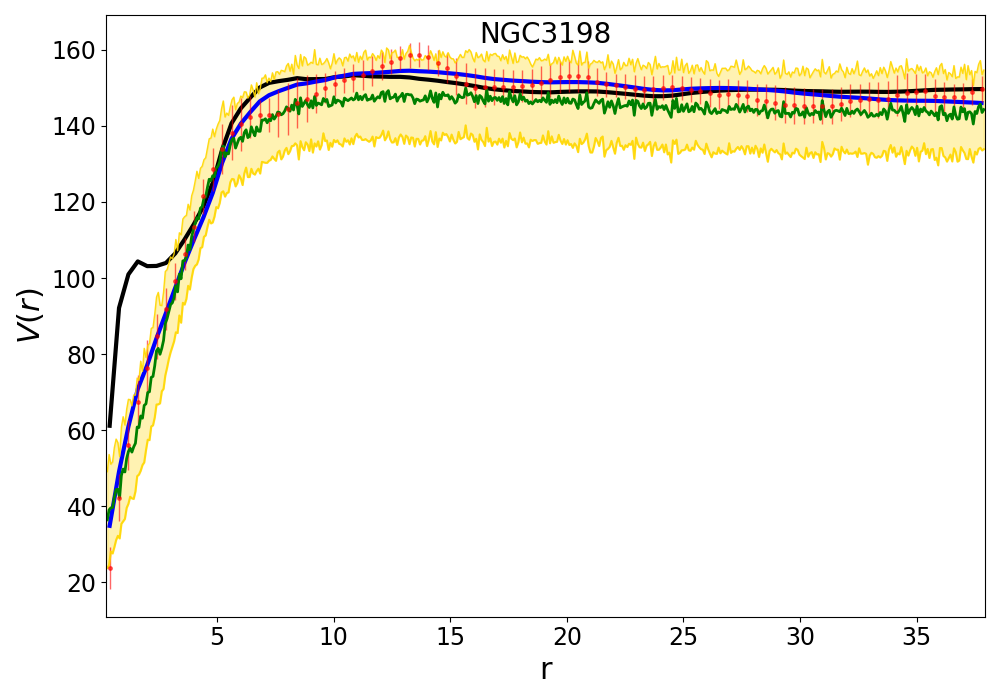} 
            \includegraphics[trim=3mm 0mm 0.0mm 0mm, clip, width=6.0cm, height=4.1cm]{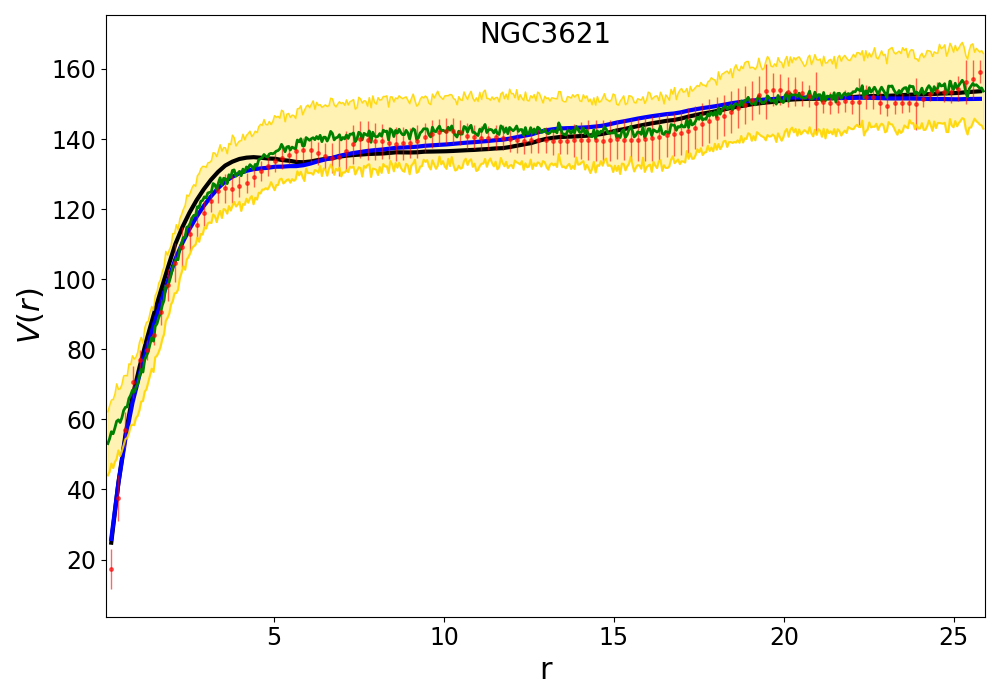} 
            \includegraphics[trim=3mm 0mm 0.0mm 0mm, clip, width=6.0cm, height=4.1cm]{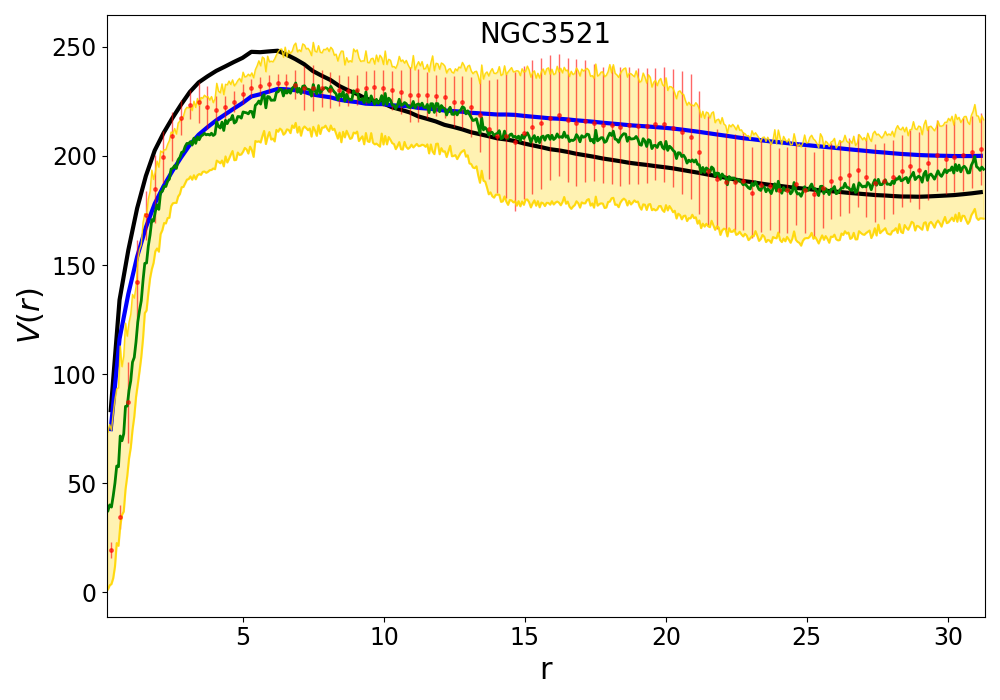}
             }
        \makebox[10.cm][c]{
            \includegraphics[trim=3mm 0mm 0.0mm 0mm, clip, width=6.0cm, height=4.1cm]{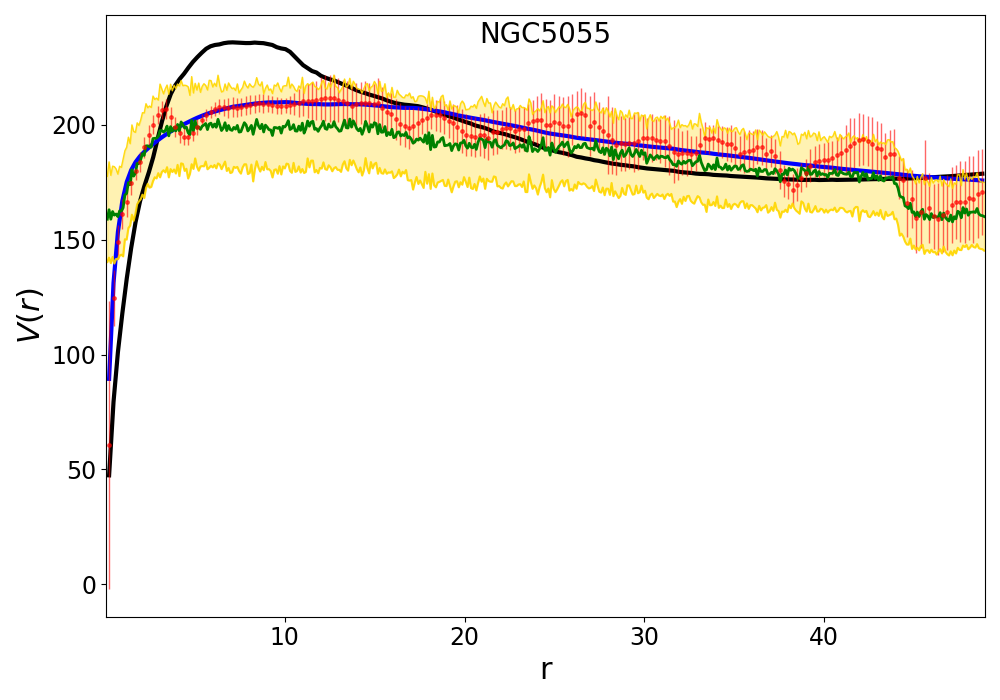} 
            \includegraphics[trim=3mm 0mm 0.0mm 0mm, clip, width=6.0cm, height=4.1cm]{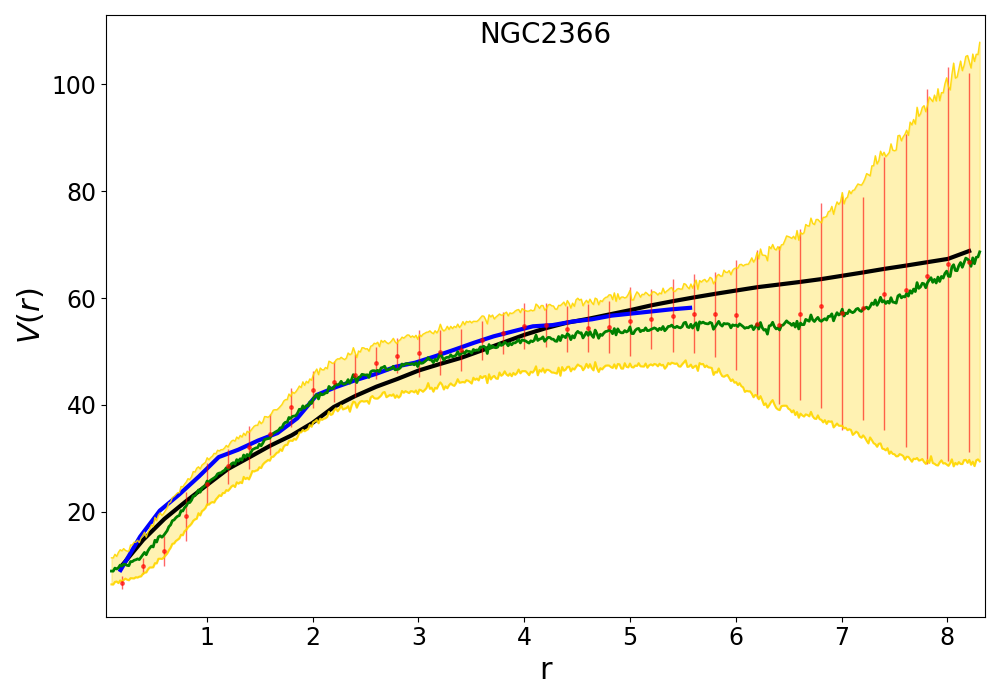} 
            \includegraphics[trim=3mm 0mm 0.0mm 0mm, clip, width=6.0cm, height=4.1cm]{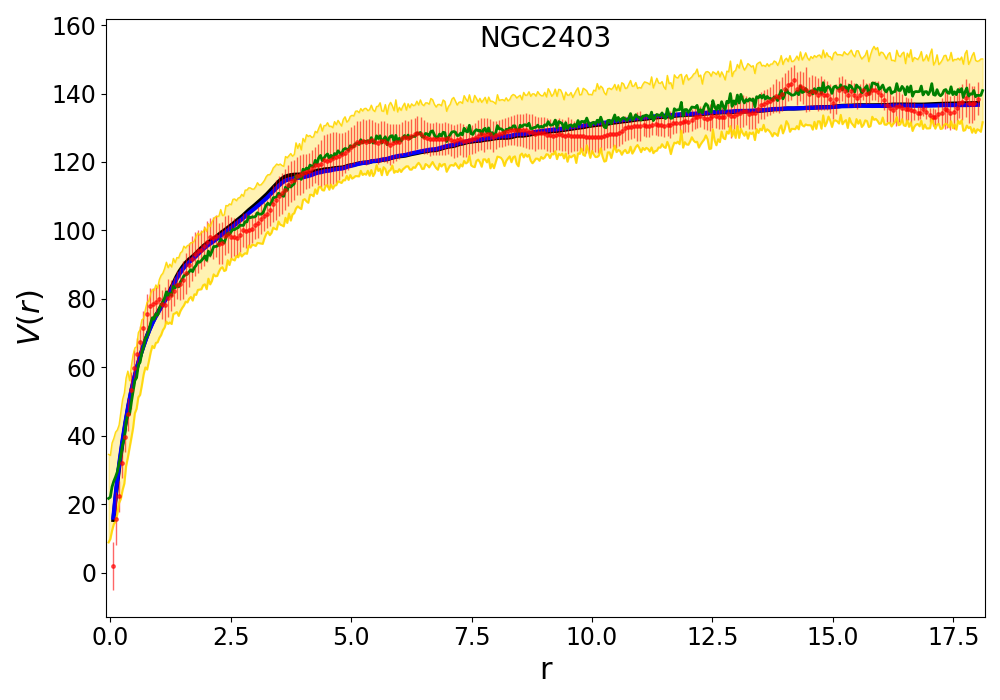}
        }
        \makebox[10.cm][c]{        
            \includegraphics[trim=3mm 0mm 0.0mm 0mm, clip, width=6.0cm, height=4.1cm]{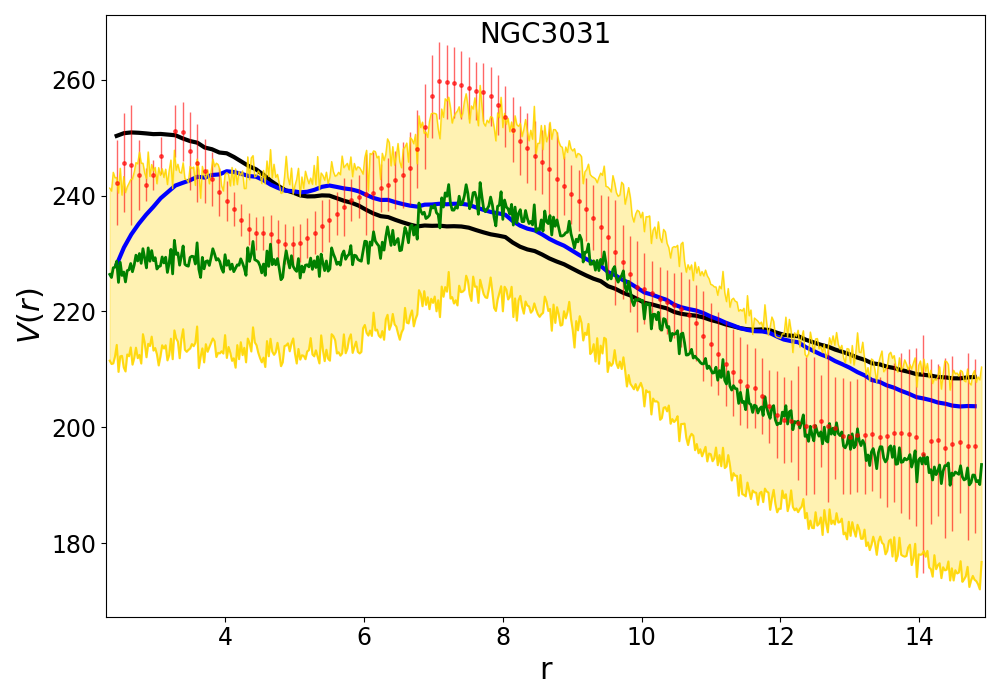}  
             \includegraphics[trim=3mm 0mm 0.0mm 0mm, clip, width=6.0cm, height=4.1cm]{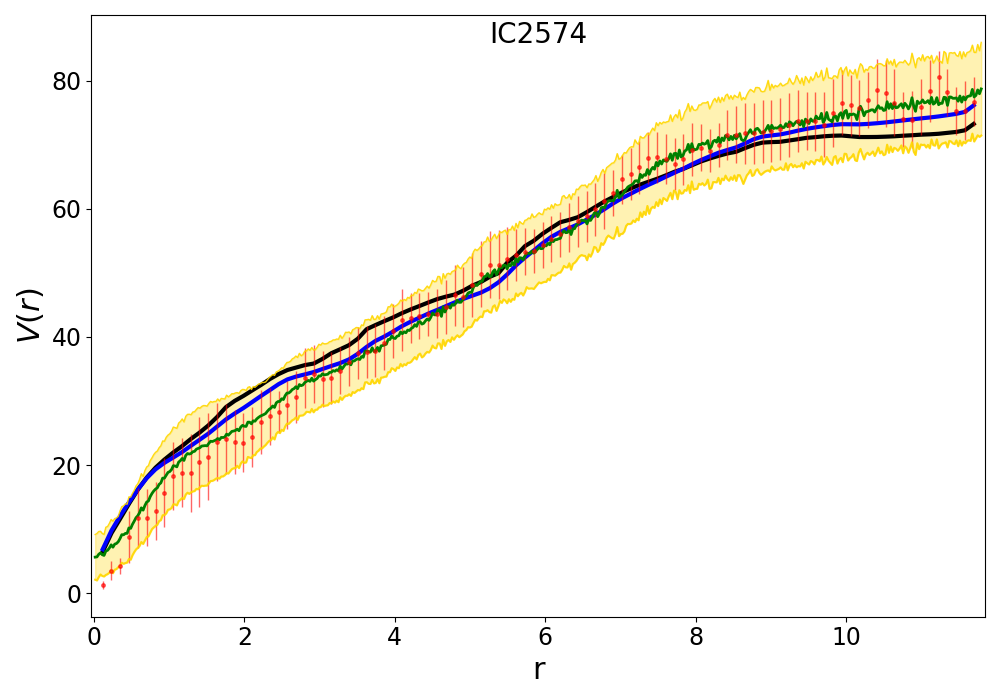}  
            }

        \caption{\footnotesize{\changes{Neural network models} for the 17 galaxies in comparison with the observational data and the NFW profiles $\Upsilon_{\star}^{free}$ and $\Upsilon_{\star}^{fix}$. The rotation velocity (y-axis) is given in km/s, whereas the radius (x-axis) is given in Kpc.}}
        \label{fig:reconstructions}
    \end{figure*}


As described above, we performed the neural network reconstructions for RCs of 17 different galaxies. Figure~\ref{fig:reconstructions} contains all the neural reconstructions for each galaxy. \changes{The green lines show the ANN model predictions for $V(r)$ (first output in Figure \ref{fig:archann}), while the yellow bands indicate the total uncertainties, given by the quadratic sum of the predicted statistical error (second output in Figure~\ref{fig:archann}) and the predictive uncertainty estimated through Monte Carlo dropout. The red points represent the observational data.}
 
To enhance our analysis, we include two well-known theoretical models, both considering an NFW dark matter halo, but the assumption regarding the mass-to-light ratio is different. In all the panels of Figure \ref{fig:reconstructions}, the black lines represent the fitting of $\Upsilon_{\star}^{\rm fix}$ as a fixed parameter, while the blue lines correspond to  $\Upsilon_{\star}^{\rm free}$ as a free parameter. The parameter estimation using these same datasets for both models was performed in \citet{deBlok:2008wp} to reproduce the total observed rotation curve.  

In Figure \ref{fig:reconstructions} we notice that the observational data points do not follow a smooth pattern. Instead, they show peaks and lows that are difficult to model accurately using parametric profiles. 
For this reason, it is interesting to generate a model based on the data \changes{that it is able to capture local variations and/or reveal regions that deviate from the overall trend,}
and perhaps make more reliable interpolations, or at least complement the predictions of the theoretical models. 

The cosmological parametric models and the neural network reconstructions are \changes{developed} independently. The parametric models have a physical basis and can be constrained using observational data, whereas the neural reconstructions are purely data-driven and do not incorporate any physical principles. Using the mean squared error as a metric, we can obtain a systematic comparison, providing insights into the relative performance of each model in capturing the observed data.

Table \ref{tab:results} summarizes the MSE values for each model and the observational data. The MSE provides a quantitative measure of how well each model (NFW free, NFW fixed, and the ANN) fits the observed rotation curves. By analyzing the MSE values, we evaluate how well the theoretical models and ANN reconstructions fit the observed data. This comparison highlights the strengths and limitations of each approach in capturing the complexity of galactic structures.

Concerning parametric models, we can notice that the MSE indicates that the one with $\Upsilon_{\star}^{\rm free}$ is closer to the data than the one with $\Upsilon_{\star}^{\rm fix}$.  This is not conclusive for preferring one model over the other, particularly because $\Upsilon_{\star}^{\rm free}$ has an additional parameter and this criterion does not penalize it.
However, an exception is noted for the galaxy DDO154, the galaxy with the lowest luminosity in the sample, where the opposite happens, although the difference is minimal, which suggests that both parametric models provide a similar overall fit to the observed data, despite differing individual contributions from DM and stars. For instance, the $\Upsilon_{\star}^{\rm free}$ model predicts a higher stellar contribution, which must be counteracted by a lower DM contribution to accurately describe the data, which results in an overestimation of the stellar luminosity of this galaxy, penalized by a higher MSE value, indicating that, unlike other galaxies, the observational $\Upsilon_{\star}$ value is particularly precise for DDO154, suggesting that no additional stellar mass is required.

  \begin{table*}
      \centering
      \captionsetup{justification=raggedright,singlelinecheck=false,font=footnotesize}
      \begin{tabular}{|c|c|c|}
        \hline
          \scriptsize{Group} & \scriptsize{Galaxies}  & \scriptsize{Criterion} \\
          \hline
           A &  \scriptsize{DDO154, NGC2903, NGC3621, NGC2403} & \scriptsize{ANNs are worst versus both NFW+$\Upsilon_{\star}$ profiles.} \\
           \hline
           B &  \scriptsize{NGC3031, NGC2841, NGC3198, NGC5055}& \scriptsize{ANNs are better than $\Upsilon_{\star}^{\rm fix}$ and worst than $\Upsilon_{\star}^{\rm free}$.} \\
           \hline
           C &  \scriptsize{NGC4736, NGC925, NGC6946, NGC7793, NGC2976, NGC7331, NGC3521, NGC2366,
IC2574} & \scriptsize{ANNs are best versus both NFW+$\Upsilon_{\star}$ profiles}. \\
           \hline
      \end{tabular}
      \caption{Classification of the galaxies considering the performance of the neural network reconstructions in comparison with the two considered NFW profiles.}
      \label{tab:groups}
  \end{table*}
On the other hand, comparing the parametric NFW+$\Upsilon_{\star}$ models with the neural reconstruction, our findings indicate that there are three different groups, summarized in Table \ref{tab:groups}. Group A comprising galaxies DDO154, NGC2903, NGC3621, and NGC2403, for which the neural reconstruction is farther from the data than both parametric models,
whereas in Group B, the MSE of the non-parametric reconstruction falls between the models, specifically, it is greater than the corresponding to $\Upsilon_{\star}^{\rm free}$ but lower than  $\Upsilon_{\star}^{\rm fix}$, the galaxies included in this group are NGC3031, NGC2841, NGC3198, and NGC5055.
Finally, in Group C, comprising the remaining nine galaxies, the non-parametric reconstruction can describe the data more closely than the parametric models used for comparison, as indicated by the lower mean squared error associated with the reconstruction. \changes{In these galaxies, the transition between stellar and dark matter dominance is not smooth, with irregular variations that make them especially difficult to capture with parametric profiles.} This suggests that for these galaxies, fitting the data well with parametric models is more difficult, indicating that there could be localized variations in the mass distribution arising from clumps of dark matter or stellar substructures.

To highlight the peculiar characteristics of the previously reported groups, we examined a galaxy from each one of them. 

\begin{itemize}
    \item The galaxy NGC3621, from Group A, exhibits an HI distribution characterized by a regularly rotating disk, with its stellar distribution modeled using a single disk. 
Both parametric models (considered in this work) and neural reconstruction effectively capture the overall shape of the rotation curve, \changes{showing the expected behavior of a regular system}. Nevertheless, discrepancies arise in the innermost parts of the galaxy, where the velocity is predominantly influenced by the stellar component. In this region, the neural reconstruction tends to over-predict the velocity, resulting in a greater deviation from the data compared to the corresponding parametric models. 
\item From Group B, we select the galaxy NGC5055, for which the parametric models suggest the existence of two exponential disks to accurately describe its stellar luminosity.
This assumption is fully satisfied when $\Upsilon_{\star}$ is fixed, resulting in two different values: the first for the inner region and the second for the outer region, which overestimates the RC in the inner region. 
However, when considering parametric models with two free $\Upsilon_{\star}$ parameters, undesirable values are obtained for the outer component, so its value is set to zero by hand.
This discrepancy explains the significant difference between \changes{the MSE of} these models. 
In contrast, the neural reconstruction captures the overall trend of the global rotation curve, \changes{ but remains slightly below the observed data across most radii, after overshooting at the very center.}   
Consequently, its MSE with the data increases in this region.
\changes{This behavior may indicate that the disk–halo decomposition in NGC5055 is not uniquely defined, since the ANN reproduces the curve without resorting to either unphysical stellar contributions or an excessively massive dark halo.}
\item From Group C, we select the galaxy NGC3521, whose mass profiles indicate that its stellar component, characterized by a single exponential disk (Eqn. \ref{eq:exponential_profile}), significantly contributes to the total distribution of matter and, consequently, to the overall RC.
Regarding parametric models, when $\Upsilon_{\star}$ is fixed, the stellar component prominently dominates and dictates the global shape of the RC, the high value of this mass-to-light ratio leads to an over-prediction in the velocity in the inner part, which is what causes the high MSE value of this model.
On the other hand, the parametric model with $\Upsilon_{\star}$ free does not over-predict the inner region; instead, it requires dark matter to fit the data at all radii. 
This causes the DM, whose contribution does not decay, to dominate the global trend of the RC at a radius of approximately 10 kpc. As a result, beyond this radius, the parametric curve remains relatively constant. However, for this model, a slight overprediction is observed at large radii, which although not as dramatic, still contributes to the high MSE of this model.
On the other hand, given the availability of data points at various distances for this galaxy, neural reconstruction achieves high accuracy in mimicking the observational data at all different radii without needing to make assumptions regarding the mass profiles. This results in a small MSE between the non-parametric reconstruction and the observational data.
\end{itemize}

A noticeable difference between the previously reported groups is that the effectiveness of the non-parametric reconstruction improves with a more evenly distributed set of data points. On closer inspection, the galaxies NGC2366 and IC2574 stand out as having the lowest mean squared error (MSE $<3$) between the neural network (ANN) predictions and the observed data. These galaxies are classified as dwarfs and are characterized by a lower stellar contribution compared to their neutral gas content, being LSB.
As for the parametric models, a discrepancy in both galaxies is that the $\Upsilon_{\star}$ free model imposes a null stellar contribution; otherwise, the predicted value becomes non-physical, because it is negative. Moreover, the ANN's superior performance in modeling these galaxies
suggests that this potential variability in $\Upsilon_{\star}$ needs to be better understood to provide a more accurate description of dark matter, which is evidently influenced by this consideration.

The \changes{groups} shown in Table \ref{tab:groups} not only highlights the flexibility of the ANN in capturing complex galactic features but also allows us to investigate the limitations of parametric models in certain cases. The results indicate that, for galaxies with more intricate mass distributions or structural features, the ANN offers a superior fit, whereas simpler systems align more closely with the NFW models
\changes{(groups A and B). For instance galaxies with the smallest MSE (fix, data) values correspond to group A, where the fixed $\Upsilon_{\star}$ model already reproduces the data accurately even if it is slightly improved by the $\Upsilon_{\star}^{free}$ assumption. } 

\section{Conclusions}
\label{sec:conclusions}
 
Our study compares the performance of \changes{theoretical and data-based neural network models} in characterizing galactic rotation curves. While theoretical parametric models, the combination of the Navarro-Frenk-White profile with $\Upsilon_{\star}$, offer detailed insights into the influence of stellar and dark matter components on galactic dynamics, neural network models stand out due to their flexibility. ANNs can adapt to the complexity of rotation curves without requiring prior assumptions about mass profiles. This comparative analysis underscores the importance of employing diverse methodological approaches to thoroughly understand galaxy dynamics in modern astrophysics.

Even with a limited number of data points for each galaxy’s rotation curve, the ANN models generate robust reconstructions that capture the essential dynamics of these systems. The results are consistent with trends suggested by traditional theoretical models, \changes{namely the rise at small radii and the nearly flat outer plateau, consistent with galaxies embedded in extended dark matter halos. At the same time, they demonstrate} that ANNs can handle sparse data while maintaining accuracy. This highlights the potential of machine learning techniques as complementary tools in astrophysical data analysis.

Modeling galaxies with a very bright bulge component and/or with non-circular motions in the inner and outer regions represents challenges for parametric approaches to accurately describe the observational data, especially because taking a $\Upsilon_{\star}$ value fully determines the profile of the velocities associated with the stellar component. This can lead to an overestimation of the velocity at the galactic center, particularly if the bulge is prominent, as indicated by the MSE calculated in Table \ref{tab:results}; in such cases, the model deviates significantly from the observed data.
In the context of parametric models, the consideration to mitigate the overestimation is to treat $\Upsilon_{\star}$ as a free parameter, avoiding maximum disk scenarios. Frequently, this modification brings the new model closer to the data.
However, as also indicated in Table \ref{tab:results}, the model may still not be closely aligned with the data, and in addition, this assumption influences the DM distribution. Also, it is important to note that in this work we are not penalizing the addition of the extra parameters.

In contrast, when the description provided by theoretical models is left out, neural reconstruction presents a promising alternative by capturing overall trends in the rotation curve, albeit with slight deviations in specific regions where the amount of data is poor, which is an expected issue given the small size of the datasets. However, neural reconstructions offer interesting alternative models for the rotation curves within the range of the radius of the observed data points.  

The success of neural reconstruction in accurately reproducing observational data is evident in Group C of galaxies, such as the rotation curve of NGC3521, which is difficult to model and where parametric models fail. \changes{In contrast, Groups A and B can still be captured by parametric models, whereas Group C highlights the galaxies where universal profiles break down and where even the transition between baryon and halo domination becomes uncertain.} This highlights the potential of machine learning techniques to improve our understanding of galactic dynamics, as they reflect the underlying astrophysical complexities and properties of the galaxies in the THINGS catalog, providing a window into the interplay between baryonic matter, dark matter, and galaxy dynamics.

\changes{The ANN framework presented in this work provides consistent and reliable modeling of galactic rotation curves, capturing the main statistical patterns of the observations without assuming any predefined functional form and relying exclusively on observational data. This approach reveals where standard parametric profiles fail and helps identify differences in the complex, non-linear dynamical behavior across galaxies. It offers a complementary, data-driven perspective for studying the interplay between baryonic and dark matter components in spiral galaxies.}


\section*{Acknowledgments}
 GGA and IGV thank the CONAHCYT postdoctoral grant and ICF-UNAM. J.A.V. acknowledges the support provided by FOSEC SEP-CONACYT Investigaci\'on B\'asica A1-S-21925, UNAM-DGAPA-PAPIIT IN117723, IN110325 and Cátedra de Investigación Marcos Moshinsky. In addition, IGV thanks the University of Geneva. The authors appreciate the help of Prof. Erwing De Blok for providing us with the data. 
\section*{Data Availability}
The data utilized in this article are sourced from the THINGS ROTATION CURVES AND MASS MODELS REPOSITORY,  \url{http://www.mpia-hd.mpg.de/THINGS/Overview.html}. These data correspond to  \citet{deBlok:2008wp} and \citet{Oh:2008ww}. For further details on the data, it is recommended to contact the authors of the original papers directly.


\appendix

\begin{figure*}
    \centering    \captionsetup{justification=raggedright,singlelinecheck=false,font=footnotesize}
    \makebox[10.cm][c]{
            \includegraphics[trim=3mm 0mm 0.0mm 0mm, clip, width=6.0cm]{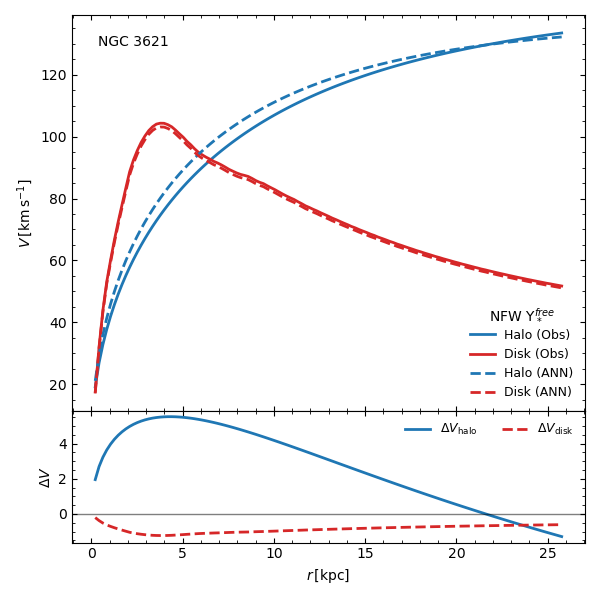}
            \includegraphics[trim=3mm 0mm 0.0mm 0mm, clip, width=6.0cm]{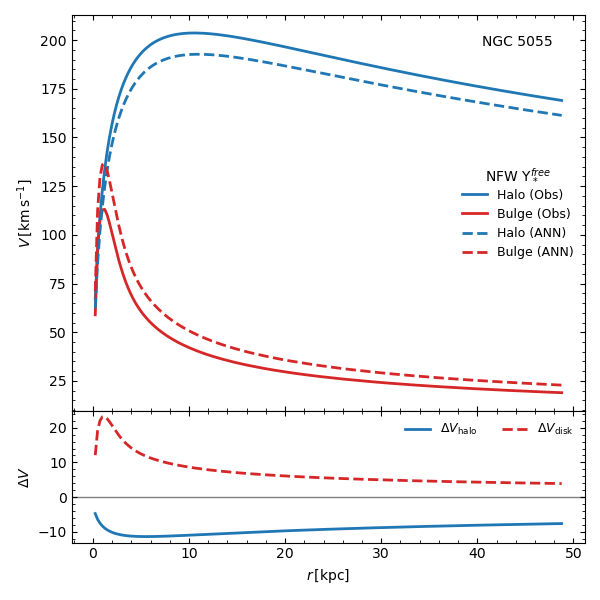}
            \includegraphics[trim=3mm 0mm 0.0mm 0mm, clip, width=6.0cm]{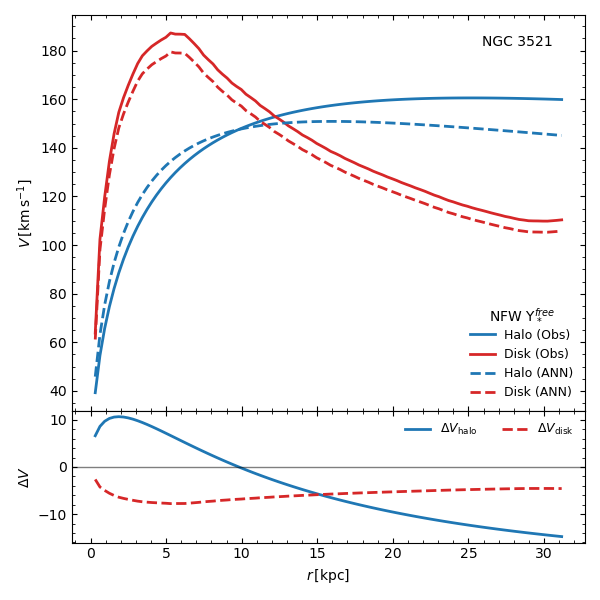}
        }
        \caption{Parametric mass contributions inferred from observed (Obs) and data-driven (ANN) rotation curves for three representative galaxies of groups A, B, and C. For each system, the upper panel shows the halo and stellar contributions obtained from independent NFW fits with free stellar mass-to-light ratio, while the lower panel displays the differences between the ANN and observation based components, $\Delta V(r) \equiv V_{\rm ANN}(r) - V_{\rm obs}(r)$. Blue curves correspond to the dark matter halo, red curves to the stellar components, while solid and dashed lines indicate fits to the observed and data-driven rotation curves, respectively.}
    \label{Fig: Ap1}
\end{figure*}

\newchanges{
\section{Parametric mass modeling of data-driven rotation curves}}
\newchanges{ In this work, we derive data-driven rotation-curve models for each galaxy, which provide predictions for both the rotation velocity and its associated uncertainty at the same radii as the observational data. This feature enables the application of the same parametric mass modeling procedure to both the observational data and the ANN predictions, using the mass decomposition described in Section \ref{sec:galaxy}.\\
In this appendix, parametric fits to both the ANN-based and observational datasets are performed within a Bayesian framework using the \texttt{emcee} sampler. The total rotation curve is modeled according to Eq. \ref{eq:total_velocity}, with a fixed gas contribution derived from the HI rotation curve, free stellar mass-to-light ratios for the disk and bulge, and a NFW dark matter halo. The goodness of fit is quantified using the reduced chi-squared statistic,
$\chi^2_\nu = \chi^2 / (N_{\rm data} - N_{\rm par})$,
where $N_{\rm data}$ is the number of data points and $N_{\rm par}$ is the number of free parameters.
\\
Figure \ref{Fig: Ap1} shows the halo and stellar contributions inferred from parametric fits to the observed and ANN-based rotation curves of three representative galaxies: NGC3621 from group A, NGC5055 from group B, and NGC3521 from group C.
For each system, the upper panel displays the contribution of each mass component to the total rotation curve, while the lower panel shows the differences between the ANN and observation based inferences.\\ 
\\
In the case of NGC3621 (left panel), the inferred stellar contributions from the observed and data-driven rotation curves are nearly identical.
The ANN-based fit, however, requires a slightly larger halo contribution, of around $\sim4\,\mathrm{km\,s^{-1}}$.
This reflects the mild overprediction of the rotation curve by the data-driven reconstruction already noted in Section~\ref{sec:results1}, which, in the parametric decomposition, is accommodated through a correspondingly higher inferred dark matter contribution. For this galaxy both datasets are well described by the parametric model, with reduced chi-squared values of $\chi^2_{\nu}=0.18$ for the ANN-based fit and $\chi^2_{\nu}=0.53$ for the observational fit.\\
\\
For NGC5055 (middle panel), allowing for free stellar mass-to-light ratios for both the disk and bulge components drives the disk contribution to formally negative values in both the observational and ANN-based fits, consistent with previous results \cite{deBlok:2008wp}. This behavior therefore does not originate from either the observational or the ANN-based rotation curves themselves. \\
As a result, the stellar contribution becomes effectively dominated by the bulge, and the disk mass-to-light ratio is set to zero in both cases.
For this galaxy, the ANN-based fit favors a larger disk contribution and a correspondingly smaller dark matter halo contribution compared to the fit obtained from the observed rotation curve. Quantitatively, both datasets are well described by the adopted parametric model, with reduced chi-squared values of $\chi^2_\nu \simeq 0.15$ for the ANN-based fit and $\chi^2_\nu \simeq 0.56$ for the observational fit.\\
\\
For NGC3521 (right panel), the inner regions of the galaxy are strongly dominated by the stellar disk in both cases.
However, the ANN-based decomposition favors a slightly reduced stellar contribution in the central region, compensated by a correspondingly larger dark matter halo contribution.
At larger radii, beyond $\sim10,\mathrm{kpc}$, the stellar contribution in the ANN-based fit remains lower than in the observational case, while the inferred halo contribution no longer is above and instead falls slightly below that obtained from the observed rotation curve.
This behavior highlights a radial redistribution between stellar and dark matter components in the ANN-based fit, rather than a uniform shift in the relative importance of the two components.
In this case, the difference between the ANN-based and observational fits is particularly pronounced, with reduced chi-squared values of $\chi^2_\nu \simeq 0.14$ and $\chi^2_\nu \simeq 5.5$, respectively.
}

\newchanges{Overall, applying the same parametric assumptions to the rotation curves reconstructed by our data-driven models shows that these curves can be consistently described by the adopted mass components, as reflected by their reduced chi-squared values. For NGC3521, this comparison highlights the sensitivity of the parametric fits to the quality of the input rotation curve.}
\bibliography{bibliography.bib}

@article{Pato:2015tja,
    author = "Pato, Miguel and Iocco, Fabio",
    title = "{The Dark Matter Profile of the Milky Way: a Non-parametric Reconstruction}",
    eprint = "1504.03317",
    archivePrefix = "arXiv",
    primaryClass = "astro-ph.GA",
    doi = "10.1088/2041-8205/803/1/L3",
    journal = "Astrophys. J. Lett.",
    volume = "803",
    number = "1",
    pages = "L3",
    year = "2015"
}

@ARTICLE{Einasto,
       author = {{Einasto}, J.},
        title = "{On the Construction of a Composite Model for the Galaxy and on the Determination of the System of Galactic Parameters}",
      journal = {Trudy Astrofizicheskogo Instituta Alma-Ata},
         year = 1965,
        month = jan,
       volume = {5},
        pages = {87-100},
       adsurl = {https://ui.adsabs.harvard.edu/abs/1965TrAlm...5...87E}
}

@ARTICLE{1914LowOB...2...66S,
       author = {{Slipher}, V.~M.},
        title = "{The detection of nebular rotation}",
      journal = {Lowell Observatory Bulletin},
         year = 1914,
        month = jan,
       volume = {2},
        pages = {66-66},
       adsurl = {https://ui.adsabs.harvard.edu/abs/1914LowOB...2...66S}
}

@article{Burkert:1995yz,
    author = "Burkert, A.",
    title = "{The Structure of dark matter halos in dwarf galaxies}",
    eprint = "astro-ph/9504041",
    archivePrefix = "arXiv",
    doi = "10.1086/309560",
    journal = "Astrophys. J. Lett.",
    volume = "447",
    pages = "L25",
    year = "1995"
}

@article{Begeman:1991iy,
    author = "Begeman, K. G. and Broeils, A. H. and Sanders, R. H.",
    title = "{Extended rotation curves of spiral galaxies: Dark haloes and modified dynamics}",
    doi = "10.1093/mnras/249.3.523",
    journal = "Mon. Not. Roy. Astron. Soc.",
    volume = "249",
    pages = "523",
    year = "1991"
}

@article{Karukes:2017kne,
    author = "Karukes, E. V. and Salucci, P.",
    title = "{The universal rotation curve of dwarf disc galaxies}",
    eprint = "1609.06903",
    archivePrefix = "arXiv",
    primaryClass = "astro-ph.GA",
    doi = "10.1093/mnras/stw3055",
    journal = "Mon. Not. Roy. Astron. Soc.",
    volume = "465",
    number = "4",
    pages = "4703--4722",
    year = "2017"
}

@article{deBlok:2008wp,
    author = "de Blok, W. J. G. and Walter, F. and Brinks, E. and Trachternach, C. and Oh, S-H. and Kennicutt, Jr., R. C.",
    title = "{High-Resolution Rotation Curves and Galaxy Mass Models from THINGS}",
    eprint = "0810.2100",
    archivePrefix = "arXiv",
    primaryClass = "astro-ph",
    doi = "10.1088/0004-6256/136/6/2648",
    journal = "Astron. J.",
    volume = "136",
    pages = "2648--2719",
    year = "2008"
}

@article{Walter:1985psa,
    author = "Walter, F. and Brinks, E. and de Blok, W. J. G. and Bigiel, F. and Kennicutt, Jr., R. C. and Thornley, M. D. and Leroy, A. K.",
    title = "{THINGS: The HI Nearby Galaxy Survey}",
    eprint = "0810.2125",
    archivePrefix = "arXiv",
    primaryClass = "astro-ph",
    doi = "10.1088/0004-6256/136/6/2563",
    journal = "Astron. J.",
    volume = "136",
    pages = "2563--2647",
    year = "2008"
}

@article{Freeman:1970mx,
    author = "Freeman, K. C.",
    title = "{On the disks of spiral and SO Galaxies}",
    doi = "10.1086/150474",
    journal = "Astrophys. J.",
    volume = "160",
    pages = "811",
    year = "1970"
}

@article{Mastache:2012ep,
    author = "Mastache, Jorge and Cervantes-Cota, Jorge L. and de la Macorra, Axel",
    title = "{Testing modified gravity at large distances with the HI Nearby Galaxy Survey\textquoteright{}s rotation curves}",
    eprint = "1212.5167",
    archivePrefix = "arXiv",
    primaryClass = "astro-ph.GA",
    doi = "10.1103/PhysRevD.87.063001",
    journal = "Phys. Rev. D",
    volume = "87",
    number = "6",
    pages = "063001",
    year = "2013"
}

@article{Navarro:1995iw,
    author = "Navarro, Julio F. and Frenk, Carlos S. and White, Simon D. M.",
    title = "{The Structure of cold dark matter halos}",
    eprint = "astro-ph/9508025",
    archivePrefix = "arXiv",
    doi = "10.1086/177173",
    journal = "Astrophys. J.",
    volume = "462",
    pages = "563--575",
    year = "1996"
}

@article{Oh:2008ww,
    author = "Oh, Se-Heon and de Blok, W. J. G. and Walter, Fabian and Brinks, Elias and Kennicutt, Jr, Robert C.",
    title = "{High-resolution dark matter density profiles of THINGS dwarf galaxies: Correcting for non-circular motions}",
    eprint = "0810.2119",
    archivePrefix = "arXiv",
    primaryClass = "astro-ph",
    doi = "10.1088/0004-6256/136/6/2761",
    journal = "Astron. J.",
    volume = "136",
    pages = "2761",
    year = "2008"
}

@article{Bell:2000jt,
    author = "Bell, Eric F. and de Jong, Roelof S.",
    title = "{Stellar mass-to-light ratios and the Tully-Fisher relation}",
    eprint = "astro-ph/0011493",
    archivePrefix = "arXiv",
    reportNumber = "ARS-2001A",
    doi = "10.1086/319728",
    journal = "Astrophys. J.",
    volume = "550",
    pages = "212--229",
    year = "2001"
}

@article{Salucci:2007tm,
    author = "Salucci, Paolo and Lapi, A. and Tonini, C. and Gentile, G. and Yegorova, I. and Klein, U.",
    title = "{The Universal Rotation Curve of Spiral Galaxies. 2. The Dark Matter Distribution out to the Virial Radius}",
    eprint = "astro-ph/0703115",
    archivePrefix = "arXiv",
    doi = "10.1111/j.1365-2966.2007.11696.x",
    journal = "Mon. Not. Roy. Astron. Soc.",
    volume = "378",
    pages = "41--47",
    year = "2007"
}

@article{Sofue:2000jx,
    author = "Sofue, Yoshiaki and Rubin, Vera",
    title = "{Rotation curves of spiral galaxies}",
    eprint = "astro-ph/0010594",
    archivePrefix = "arXiv",
    reportNumber = "U-TOKYO-ASTRO-PREPRINT-2000-09, U-Tokyo Astro. Preprint No. 2000-09",
    doi = "10.1146/annurev.astro.39.1.137",
    journal = "Ann. Rev. Astron. Astrophys.",
    volume = "39",
    pages = "137--174",
    year = "2001"
}

@article{Rubin:1970zza,
    author = "Rubin, Vera C. and Ford, Jr., W. Kent",
    title = "{Rotation of the Andromeda Nebula from a Spectroscopic Survey of Emission Regions}",
    doi = "10.1086/150317",
    journal = "Astrophys. J.",
    volume = "159",
    pages = "379--403",
    year = "1970"
}

@article{Pahwa_2018,
   title={Structural properties of faint low-surface-brightness galaxies},
   volume={478},
   ISSN={1365-2966},
   url={http://dx.doi.org/10.1093/mnras/sty1139},
   DOI={10.1093/mnras/sty1139},
   number={4},
   journal={Monthly Notices of the Royal Astronomical Society},
   publisher={Oxford University Press (OUP)},
   author={Pahwa, Isha and Saha, Kanak},
   year={2018},
   month=may, pages={4657–4668} }

@article{Navarro-Boullosa:2023bya,
    author = "Navarro-Boullosa, Atalia and Bernal, Argelia and Vazquez, J. Alberto",
    title = "{Bayesian analysis for rotational curves with \ensuremath{\ell}-boson stars as a dark matter component}",
    eprint = "2305.01127",
    archivePrefix = "arXiv",
    primaryClass = "astro-ph.CO",
    doi = "10.1088/1475-7516/2023/09/031",
    journal = "JCAP",
    volume = "09",
    pages = "031",
    year = "2023"
}

@article{deBlok:1997zlw,
    author = "de Blok, W. J. G. and McGaugh, S. S.",
    title = "{The Dark and visible matter content of low surface brightness disk galaxies}",
    eprint = "astro-ph/9704274",
    archivePrefix = "arXiv",
    doi = "10.1093/mnras/290.3.533",
    journal = "Mon. Not. Roy. Astron. Soc.",
    volume = "290",
    pages = "533--552",
    year = "1997"
}

@article{Sellwood:1998kj,
    author = "Sellwood, Jerry A. and Moore, E. M.",
    title = "{On the formation of disk galaxies and massive central objects}",
    eprint = "astro-ph/9807010",
    archivePrefix = "arXiv",
    reportNumber = "RUTGERS-ASTROPHYSICS-PREPRINT-230",
    doi = "10.1086/306557",
    journal = "Astrophys. J.",
    volume = "510",
    pages = "125--135",
    year = "1999"
}

@article{Kennicutt:2003dc,
    author = "Kennicutt, Jr., R. C. and others",
    title = "{SINGS: The SIRTF Nearby Galaxies Survey}",
    eprint = "astro-ph/0305437",
    archivePrefix = "arXiv",
    doi = "10.1086/376941",
    journal = "Publ. Astron. Soc. Pac.",
    volume = "115",
    pages = "928--952",
    year = "2003"
}

@article{fernandez2019galaxy,
  title={Galaxy rotation curves using a non-parametric regression method: core, cuspy and fuzzy scalar field dark matter models},
  author={Fern{\'a}ndez-Hern{\'a}ndez, Lizbeth M and Montiel, Ariadna and Rodr{\'\i}guez-Meza, Mario A},
  journal={Monthly Notices of the Royal Astronomical Society},
  volume={488},
  number={4},
  pages={5127--5144},
  year={2019},
  publisher={Oxford University Press}
}

@article{deBlok:2002vgq,
    author = "de Blok, W. J. G. and Bosma, A.",
    title = "{High-resolution rotation curves of low surface brightness galaxies}",
    eprint = "astro-ph/0201276",
    archivePrefix = "arXiv",
    doi = "10.1051/0004-6361:20020080",
    journal = "Astron. Astrophys.",
    volume = "385",
    pages = "816",
    year = "2002"
}

@article{hornik1990universal,
  title={Universal approximation of an unknown mapping and its derivatives using multilayer feedforward networks},
  author={Hornik, Kurt and Stinchcombe, Maxwell and White, Halbert},
  journal={Neural networks},
  volume={3},
  number={5},
  pages={551--560},
  year={1990},
  publisher={Elsevier},
  doi={https://doi.org/10.1016/0893-6080(90)90005-6}
}

@article{gal2015dropout,
  title={Dropout as a bayesian approximation: Insights and applications},
  author={Gal, Yarin and Ghahramani, Zoubin},
  journal={Deep Learning Workshop, ICML},
  volume={1},
  pages={2},
  year={2015},
}

@article{gomez2023neuralrecs,
  title={Neural network reconstructions for the Hubble parameter, growth rate and distance modulus},
  author={G{\'o}mez-Vargas, Isidro and Medel-Esquivel, Ricardo and Garc{\'\i}a-Salcedo, Ricardo and V{\'a}zquez, J Alberto},
  journal={The European Physical Journal C},
  volume={83},
  number={4},
  pages={304},
  year={2023},
  publisher={Springer}
}

@article{gomez2023neuralgenetic,
  title={Neural networks optimized by genetic algorithms in cosmology},
  author={G{\'o}mez-Vargas, Isidro and Andrade, Joshua Briones and V{\'a}zquez, J Alberto},
  journal={Physical Review D},
  volume={107},
  number={4},
  pages={043509},
  year={2023},
  publisher={APS}
}

@article{mitragomez2024dark,
  title={Dark energy reconstruction analysis with artificial neural networks: Application on simulated Supernova Ia data from Rubin Observatory},
  author={Mitra, Ayan and G{\'o}mez-Vargas, Isidro and Zarikas, Vasilios},
  journal={Physics of the Dark Universe},
  pages={101706},
  year={2024},
  publisher={Elsevier}
}

@book{goodfellow2016deep,
  title={Deep learning},
  author={Goodfellow, Ian and Bengio, Yoshua and Courville, Aaron and Bengio, Yoshua},
  volume={1},
  series={},
  year={2016},
  publisher={MIT press Cambridge}
}

@book{nielsen2015neural,
  title={Neural networks and deep learning},
  author={Nielsen, Michael A},
  volume={25},
  series={},
  year={2015},
  publisher={Determination press San Francisco, CA, USA}
}

@article{DEAP_JMLR2012, 
    author    = " F\'elix-Antoine Fortin and Fran\c{c}ois-Michel {De Rainville} and Marc-Andr\'e Gardner and Marc Parizeau and Christian Gagn\'e ",
    title     = { {DEAP}: Evolutionary Algorithms Made Easy },
    pages    = { 2171--2175 },
    volume    = { 13 },
    month     = { 07},
    year      = { 2012 },
    journal   = { Journal of Machine Learning Research }
}

@article{benatan2023enhancing,
  title={{Enhancing Deep Learning with Bayesian Inference}: {Create More Powerful, Robust Deep Learning Systems with Bayesian Deep Learning in Python}},
  author={Benatan, Matt and Gietema, Jochem and Schneider, Marian},
  year={2023},
  journal={Packt Publishing}
}

@article{arguelles2023galaxy,
  title={Galaxy Rotation Curve Fitting Using Machine Learning Tools},
  author={Arg{\"u}elles, Carlos R and Collazo, Santiago},
  journal={Universe},
  volume={9},
  number={8},
  pages={372},
  year={2023},
  publisher={MDPI}
}

@article{tanoglidis2021deepshadows,
  title={DeepShadows: Separating low surface brightness galaxies from artifacts using deep learning},
  author={Tanoglidis, Dimitrios and {\'C}iprijanovi{\'c}, Aleksandra and Drlica-Wagner, Alex},
  journal={Astronomy and Computing},
  volume={35},
  pages={100469},
  year={2021},
  publisher={Elsevier}
}

@article{thuruthipilly2024shedding,
  title={Shedding light on low-surface-brightness galaxies in dark energy surveys with transformer models},
  author={Thuruthipilly, H and Junais, Junais and Pollo, A and Sureshkumar, U and Grespan, M and Sawant, P and Ma{\l}ek, K and Zadrozny, A},
  journal={Astronomy \& Astrophysics},
  volume={682},
  pages={A4},
  year={2024},
  publisher={EDP Sciences}
}

@article{yi2022automatic,
  title={Automatic detection of low surface brightness galaxies from Sloan Digital Sky Survey images},
  author={Yi, Zhenping and Li, Jia and Du, Wei and Liu, Meng and Liang, Zengxu and Xing, Yongguang and Pan, Jingchang and Bu, Yude and Kong, Xiaoming and Wu, Hong},
  journal={Monthly Notices of the Royal Astronomical Society},
  volume={513},
  number={3},
  pages={3972--3981},
  year={2022},
  publisher={Oxford University Press}
}

@article{cortes2024galaxies,
  title={Galaxies in the Zone of Avoidance: Misclassifications using machine learning tools},
  author={Cort{\'e}s, P Marchant and Castell{\'o}n, JL and Alonso, MV and Baravalle, L and Villal{\'o}n, C and Sgr{\'o}, MA and Daza-Perilla, IV and Soto, M and Castro, F Milla and Minniti, D and others},
  journal={arXiv preprint arXiv:2403.03098},
  year={2024}
}

@article{wang2020reconstructing,
  title={Reconstructing Functions and Estimating Parameters with Artificial Neural Networks: A Test with a Hubble Parameter and SNe Ia},
  author={Wang, Guo-Jian and Ma, Xiao-Jiao and Li, Si-Yao and Xia, Jun-Qing},
  journal={Astrophysical Journal Supplement Series},
  volume={246},
  number={1},
  pages={13},
  year={2020},
  publisher={IOP Publishing},
  doi={https://doi.org/10.3847/1538-4365/ab620b}
}

@article{l2020defying,
  title={Defying the laws of gravity I: model-independent reconstruction of the Universe expansion from growth data},
  author={L’Huillier, Benjamin and Shafieloo, Arman and Polarski, David and Starobinsky, Alexei A},
  journal={Monthly Notices of the Royal Astronomical Society},
  volume={494},
  number={1},
  pages={819--826},
  year={2020},
  publisher={Oxford University Press},
  doi={https://doi.org/10.1093/mnras/staa633}
}

@article{escamilla2023model,
  title={Model-independent reconstruction of the interacting dark energy kernel: Binned and Gaussian process},
  author={Escamilla, Luis A and Akarsu, {\"O}zg{\"u}r and Di Valentino, Eleonora and Vazquez, J Alberto},
  journal={Journal of Cosmology and Astroparticle Physics},
  volume={2023},
  number={11},
  pages={051},
  year={2023},
  publisher={IOP Publishing}
}

@article{Keeley:2020aym,
	doi = {10.3847/1538-3881/abdd2a},
	url = {https://doi.org/10.3847/1538-3881/abdd2a},
	month = {Feb},
	publisher = {American Astronomical Society},
	volume = {161},
	number = {3},
	pages = {151},
	author = {Ryan E. Keeley and Arman Shafieloo and Gong-Bo Zhao and Jose Alberto Vazquez and Hanwool Koo},
	title = {Reconstructing the Universe: Testing the Mutual Consistency of the Pantheon and {SDSS}/{eBOSS} {BAO} Data Sets with Gaussian Processes},
	journal = {The Astronomical Journal},
    year = "2021"
}

@article{sharma2020reconstruction,
  title={Reconstruction of late-time cosmology using Principal Component Analysis},
  author={Sharma, Ranbir and Mukherjee, Ankan and Jassal, HK},
  journal={preprint arXiv:2004.01393},
  year={2020}
}

@article{kingma2014adam,
  title={Adam: A method for stochastic optimization},
  author={Kingma, Diederik P and Ba, Jimmy},
  journal={arXiv preprint arXiv:1412.6980},
  year={2014}
}

@article{Escamilla:2021uoj,
    author = "Escamilla, Luis A. and Vazquez, J. Alberto",
    title = "{Model selection applied to reconstructions of the Dark Energy}",
    eprint = "2111.10457",
    archivePrefix = "arXiv",
    primaryClass = "astro-ph.CO",
    doi = "10.1140/epjc/s10052-023-11404-2",
    journal = "Eur. Phys. J. C",
    volume = "83",
    number = "3",
    pages = "251",
    year = "2023"
}

@article{Velazquez:2024aya,
    author = "Vel{\'a}zquez, Jos{\'e} de Jes{\'u}s and Escamilla, Luis A. and Mukherjee, Purba and V{\'a}zquez, J. Alberto",
    title = "{Non-Parametric Reconstruction of Cosmological Observables Using Gaussian Processes Regression}",
    eprint = "2410.02061",
    archivePrefix = "arXiv",
    primaryClass = "astro-ph.CO",
    doi = "10.3390/universe10120464",
    journal = "Universe",
    volume = "10",
    number = "12",
    pages = "464",
    year = "2024"
}
\end{document}